\documentclass[10pt]{article}

\usepackage[numbers,square]{natbib}

\usepackage{xcolor}
\usepackage{epsfig}
\usepackage{cancel}
\usepackage{caption}
\usepackage{feynmf} 
\usepackage{amssymb}
\usepackage{amsfonts}
\usepackage{epsf}
\usepackage{rotating}
\usepackage{graphicx}
\usepackage{amsmath}
\usepackage{fancyhdr}
\usepackage{subfigure}
\usepackage{graphics}
\usepackage{pstricks}
\usepackage{color}
\usepackage{frontespizio}
\usepackage{hyperref}
\hypersetup{
	colorlinks,
	citecolor=green,
	filecolor=black,
	linkcolor=blue,
	urlcolor=black
}
\usepackage{type1ec}
\usepackage[T1]{fontenc}
\usepackage{lettrine}
\usepackage{bbold}
\usepackage{calligra}
\usepackage{tikz}
\usepackage{subfigure}
\usepackage{mathrsfs}
\usepackage{curve2e}
\usepackage{setspace}
\usepackage{indentfirst}
\usepackage{emptypage}
\usepackage[babel]{csquotes} %bibliografia
\usepackage[font=small,labelfont=bf,labelsep=quad]{caption} %tabelle e figure
\usepackage{graphicx} %figure
\usepackage{listings} %codici
\usetikzlibrary{patterns}
\usepackage{relsize}
\usetikzlibrary{intersections,positioning}
\usetikzlibrary{decorations.pathmorphing,decorations.markings,arrows,positioning}
\usepackage{braket}
\usepackage{mathrsfs}
\usepackage{stackengine}
\usepackage{calc}
\newlength\shlength
\newcommand\xshlongvec[2][0]{\setlength\shlength{#1pt}%
	\stackengine{-5.6pt}{$#2$}{\smash{$\kern\shlength%
			\stackengine{7.55pt}{$\mathchar"017E$}%
			{\rule{\widthof{$#2$}}{.57pt}\kern.4pt}{O}{r}{F}{F}{L}\kern-\shlength$}}%
	{O}{c}{F}{T}{S}}

\IfFileExists{dsfont.sty}
{\usepackage{dsfont}
	\let\mathbb=\mathds
	\newcommand{\id}{\mathds{1}}}
{\typeout{Package dsfont.sty was not found, using alternative macros.}
	\let\mathds=\mathbb
	\newcommand{\id}{\mbox{1 \kern-.59em {\rm l}}}}

\usepackage{slashed}
\usepackage{units}
\usepackage{setspace}
\def\nbox#1#2{\vcenter{\hrule \hbox{\vrule height#2in
			\kern#1in \vrule} \hrule}}
\def\sq{\,\raise.5pt\hbox{$\nbox{.09}{.09}$}\,}
\def\sqb{\,\raise.5pt\hbox{$\overline{\nbox{.09}{.09}}$}\,}

\newcommand{\bea}{\begin{eqnarray}}
\newcommand{\eea}{\end{eqnarray}}
\newcommand{\be}{\begin{equation}}
\newcommand{\ee}{\end{equation}}

\newcommand{\bes}{\begin{subequations}}
	\newcommand{\ees}{\end{subequations}}

\numberwithin{equation}{section}

\usepackage{accents}

\makeatletter
\newcommand{\xLine}[2][]{\ext@arrow 0359\Rightarrowfill@{#1}{#2}}
\makeatother
\xdefinecolor{darkgreen}{RGB}{102, 204, 70}
\xdefinecolor{darkblue}{RGB}{0, 0, 153}

\begin{document}

\begin{center}

{\bf \Large Conformal Unification 
in a Quiver Theory\\ and Gravitational Waves\\
}
\vspace{2.0cm}
{\bf Claudio Corian\`{o}, Paul H. Frampton and Alessandro Tatullo\\ }
\vspace{0.5cm}
{\it INFN Sezione d Lecce,\\
 Dipartimento di Matematica e Fisica,\\ 
Universit\`{a} del Salento,\\ Via Arnesano, 73100 Lecce, Italy\footnote{claudio.coriano@le.infn.it, paul.h.frampton@gmail.com, alessandro.tatullo@le.infn.it}
}
\end{center}
\begin{abstract}
\noindent
The detection of a stochastic background of gravitational waves can reveal
details about first-order phase transitions (FOPTs) at a time 
of $10^{-13}s$ of the early
universe. We specifically discuss quiver-type GUTs which avoid both
proton decay and a desert hypothesis. A quiver based on $SU(3)^{12}$
which breaks at a $E=4000$ GeV to trinification $SU(3)^3$ has
a much larger ($g_*=1,272$) number of effective massless degrees of freedom
than the Standard Model. Assuming a FOPT for this model, we investigate the strain sensitivity of 
typical of this model for a wide range of FOPT parameters.
\end{abstract}
\newpage
\section{Introduction} 

\noindent
Since the discovery of gravitational waves from the merger of two black
holes, each with mass $M_{BH}\sim 30$ $M_{\odot}$, announced as event
GW150914 in 2016\cite{Abbott:2016blz} by the LIGO-Virgo Collaboration,
it has become clear that this provides a new and invaluable window into
the early universe. Many subsequent similar observations have occurred
and of special interest is one where two neutron stars merger \cite{GBM:2017lvd}
where the event was shortly thereafter observed electromagnetically, thereby
giving birth to multi-messenger astronomy.

\noindent
The conventional way of seeking new physics at the highest possible energies is by particle colliders, with the highest energy of any active collider is at the LHC (Large Hadron Collider) with center of mass (com) energy 14 TeV. Possible colliders with 
center of mass energies up to
100 TeV are under discussion. In the early universe, such energy / temperature
existed at cosmic times with $t < 10^{-16}$ s. To study higher energies or
shorter cosmic times a method may be provided by gravitational wave detectors(GWDs) which can, in
principle, be sensitive to signals generated from all cosmic times back to the
Planck time $t \sim 10^{-44}$ s, which could lay bare 14 more orders of magnitude
in energies up to the Planck, $M_{Planck} \sim 10^{19}$ GeV \cite{Binetruy:2012ze}.\\
In the present article, we more conservatively study energies up to a few TeV
which may overlap with accessible collider energies, yet where the detection of GWs could
give additional information about the type of phase transitions which occurred
in the early universe. The discovery in \cite{Abbott:2016blz} has already precipitated
a number of papers 
(see, for instance, \cite{Niksa:2018ofa,Geller:2018mwu,Wan:2018udw,Megias:2018sxv,Hindmarsh:2019phv,Alanne:2019bsm,Ellis:2020awk,DeCurtis:2019rxl,DelleRose:2019pgi}) 
which discuss this possibility. This process takes off from the analysis of binary mergers, which can shed light on the quark-hadron phase transition \cite{Weih:2019xvw}, to far larger scales. \\
Although such experiments could eventually investigate phase transitions
up to the GUT scale {\it e.g.} $10^{16}$ GeV, the earliest such linkage is likely
to come at a much lower energy. \\
The advent of the AdS/CFT correspondence \cite{Maldacena:1997re} 
between a maximally supersymmetric ${\cal N}=4$ gauge theory with gauge group
$SU(N)$, in a limit where $N \rightarrow \infty$, and a Type IIB superstring theory compactified on a manifold
$AdS_5 \times S^5$, has introduced a hitherto unexpected connection between the two interactions. This has provided a number of insights
into solution of problems in a broad range of theoretical physics.\\
To make a connection to particle phenomenology, it was then proposed
that a generalisation of \cite{Maldacena:1997re}, which broke supersymmetry
completely from ${\cal N}=4$ to ${\cal N}=0$ with finite N, should be considered.
This was attained by using a generalised manifold $AdS_5 \times S^5/Z_p$,
an orbifold, leading to a gauge group $SU(N)^p$ and matter fields most
conveniently characterised as bifundamental and adjoint representations 
in a quiver diagram; hence the name quiver theory
\cite{Frampton:1998en,Frampton:1983nr,Frampton:1983ez,Frampton:2011rh}.\\
One especially interesting example was discussed over a decade later 
\cite{Frampton:2002st,Frampton:2003cp}. It uses the values $p=12$ and $N=3$ and gives
rise to a theory which unifies at an unusually low energy scale 
$E\simeq4$ TeV. Proton decay is absent due to the quiver construction. The goal of this work is to introduce the analysis of such models in a preliminary way, trying to uncover their possible impact on future GW research.\\
Given the significant interest  in the detection of stochastic GWs, relics of the early universe, it is forseeable that such alternative scenarios to ordinary GUTs may draw the attention of new experimental proposals in the near future by LIGO \cite{TheLIGOScientific:2014jea}, ET \cite{Punturo:2010zz}\cite{Maggiore:2019uih}, MAGIS \cite{Graham:2017pmn}, AEDGE \cite{Bertoldi:2019tck} and LISA 
\cite{Audley:2017drz} \cite{Barausse:2020rsu}.

\section{The Quiver Model}

We use a different strategy for unification of electroweak theory with QCD than in
GUTs based on $SU(5)$ or $SO(10)$. The choice of quiver is motivated by bottom-up
considerations. The desert with logarithmic running of couplings is abandoned. Instead, 
the standard $SU(3)_C \times SU(2)_L \times U(1)_Y$ gauge group is embedded 
in a semi-simple gauge group such as $SU(3)^p$ as suggested by gauge theories arising from compactification of the IIB superstring on an orbifold $AdS_5 \times S^5/\Gamma$ 
where $\Gamma$ is the abelian finite group $Z_p$. In such non-supersymmetric 
quiver gauge theories the unification of couplings occurs abruptly at $\mu = M$ through 
the diagonal embeddings of 321 in $SU(3)^p$. The key prediction of such 
unification shifts from proton decay to additional particle content, 
in the present model at $\sim 4$ TeV.
We use the RG $\beta$-functions from \cite{Amaldi:1991zx}.
Taking the values at the $Z$-pole $\alpha_Y (M_Z) = 0.0101$, $\alpha_2(M_Z ) = 0.0338$ and $\alpha_3(M_Z ) = 0.118$, they are taken to run between $M_Z$ and $M$ according to the SM equations
\begin{eqnarray}
\alpha_Y(M)&=& (0.01014)^{-1} - (41/12\pi)\ln(M/M_Z) \nonumber \\
                   &=& 98.619 - 1.0876 y \\
\alpha^{-1}(M) & = & 0.0338)^{-1}+(19/12\pi)\ln(M/M_Z) \nonumber  \\
&=& 29.586 + 0.504y
\end{eqnarray}
\begin{eqnarray}
\alpha^{-1}(M) &=& (0.118)^{-1} + (7/2\pi)\ln(M/M_Z) \nonumber \\
&=& 8.474 + 1.114y  \end{eqnarray}

\noindent
where $y = \log(M/M_Z)$.

\noindent
The scale at which 
\begin{equation}
\sin^2\theta(M) = \alpha_Y (M)/(\alpha_2(M)+ \alpha_Y (M))
\end{equation}
satisfies $\sin^2\theta(M) = 1/4$ is at a value $M \simeq 4$ TeV.\\
\noindent
We now focus on the ratio 
\begin{equation}
R(M) \equiv \alpha_3(M)/\alpha_2(M).  \end{equation}
We find that 
\begin{equation}
R(M_Z) \simeq
 3.5, \qquad R(M_3) = 3, \qquad R(M_{5/2}) = 5/2,\qquad R(M_2) = 2
 \end{equation}
occur at the scales 
\begin{equation}
M_3 \simeq 400\, \textrm{GeV}, \qquad M_{5/2} \simeq 4\,\textrm{TeV}  \qquad \textrm{and}\qquad M_2 = 140\,\textrm{TeV}. 
\end{equation}

\noindent
The proximity of $M_{5/2}$ and $M$, accurate to a few percent, suggests strong-electroweak 
unification at $\sim 4$ TeV.
There remains the question of embedding such unification in an $SU(3)^p$ of the quiver type discussed in the Introduction.\\
 Since the required embeddings
of $SU(2)_L \times U(1)_Y$
into an $SU(3)$ necessitates $3\alpha_Y = \alpha_H$, the ratios of couplings at $\simeq 4$ TeV is
\begin{equation}
\alpha_{3C} : \alpha_{3W} : \alpha_{3H} :: 5:2:2
\end{equation}
and thus it is natural to examine $p=12$
with diagonal embeddings of Colour (C), Weak (W) and Hypercharge (H)
in $SU(3)^2,SU(3)^5,SU(3)^5$,  respectively.

To accomplish this we specify the embedding of $\Gamma = Z_{12}$ 
in the global $SU(4)$ R-parity of the ${\cal N} = 4$ supersymmetry 
of the underlying theory.\\
 Defining $\alpha = \exp(2\pi i/12)$, 
this specification can be made by 
\begin{equation}
{\bf 4} \equiv (\alpha^{A_1},\alpha^{A_2},\alpha^{A_3},\alpha^{A_4})\,\,\,
\textrm{with} \,\, \sum A_{\mu}  = 0\,\,\, (\textrm{mod} \,\,\,12) 
\end{equation}
and all $A_{\mu} \neq 0$ so that all four supersymmetries are broken  from ${\cal N} =4$ to ${\cal N} =0$.

\noindent
Having specified $A_{\mu}$ we calculate the content of complex scalars by investigating in $SU(4)$ the 
\begin{equation}
{\bf 6} 
\equiv (\alpha^{a_1},\alpha^{a_2},\alpha^{a_3},\alpha^{-a_3},\alpha^{-a_2},\alpha^{-a_1}) \end{equation}
with
\begin{equation}
 a_1 = A_1 + A_2,\,\,\,\, a_2=A_2+A_3,\,\,\,\,  a_3=A_3+A_1  \,\,\,\,(\textrm{mod}\,\, 12)
 \end{equation}
 where all
quantities are defined (mod 12).
Finally we identify the nodes as C, W or H on the dodecahedral quiver such that the complex scalars

\begin{equation}
\sum_{i=1}^{3}\sum_{\alpha=1}^{12} \left(N_{\alpha} ,\bar{N}_{\alpha+a_{i}} \right) 
\end{equation}
are adequate to allow the required symmetry breaking to the $SU(3)^3$ diagonal subgroup,
and the chiral fermions given by

\begin{equation}
\sum_{\mu=1}^{4}\sum_{\alpha=1}^{12} \left(N_{\alpha}, \bar{N}_{\alpha+A_{\mu}} \right)
\end{equation}
will be able to include the thee generations of fermions. These constraints
are nontrivial but a solution was provided in \cite{Frampton:2002st}.

\noindent
The unique solution is to adopt $A_{\mu} \equiv (1,2,3,6)$ and for the quiver nodes
take the ordering:

\begin{equation}
- C - W - H- C - W^4 - H^4 -
\label{quiver}
\end{equation}
with the two ends of Eq.(\ref{quiver}) identified to form a dodecahedral quiver.

\noindent
With this choice the scalars are provided by $A_I = (3,4,5)$ and are sufficient
to break all the diagonal subgroups to

\begin{equation}
SU(3)_C \times SU(3)_W \times SU(3)_H
\label{trinification}
\end{equation}
and the choice of quiver nodes in Eq. (\ref{quiver}) generates precisely three
quark lepton families which transform under Eq.(\ref{trinification}) as

\begin{equation}
3 \left[ \left( 3, \bar{3}, 1 \right) + \left( 1, 3, \bar{3} \right) + \left( \bar{3}, 1, 3 \right) \right]
\label{families}
\end{equation}

\noindent
The ordering of the quiver nodes in Eq.(\ref{quiver}) merits further explication. \\
The
point is that breaking to a diagonal subgroup $SU(3)$ from $SU(3)^r$ is possible
if and only if all the $r$ nodes are connected by bifundamental scalars and no node
is isolated. By trial and error, the reader can become convinced that Eq.(\ref{quiver})
is the unique choice which satisfies this highly restrictive constraint.\\
 Once the number of C, W and H nodes has been chosen in order that the three couplings
accurately unify, there is generally no quiver diagram which will allow the
required symmetry breaking. \\
We have found only very few successful
examples, one of which is studied assiduously in this article. The choice
of gauge group and matter fields is far less arbitrary than it may seem at
first sight. The choice is unique, or nearly unique.\\
Anomaly freedom of the superstring guarantees that the only
possible combination of chiral fermions is as in Eq. (\ref{families}).
This fact makes it easier to confirm the occurrence of three familes
in the complicated quiver diagram because one needs to check
only one of the three representations, for example the colour
triplets which all originate from C nodes.\\
Further breaking to the SM group gives the correct light chiral states.
The couplings run up to $E=M$ and then become frozen for at least
a finite energy range provided conformal invariance sets in as expected
by analogy with the supersymmetric case in \cite{Maldacena:1997re}.\\
At $M\sim 4$TeV, there are many new particles predicted by this scenario:
gauge bosons, fermions and scalars. These are necessary
to satisfy the conformal constraints discussed in \cite{Frampton:1998en}.\\
This quiver model is interesting because it ameliorates the hierarchy
problem in $SU(5)$ and $SO(10)$ GUTs between the weak / Higgs mass
scale and the GUT scale. It predicts correctly the value of
$\sin^2 \theta (M_Z)$, of $\alpha_3 (M_Z)$ and the appearance
of exactly three families.\\
One final advantage is that the unification of the three SM couplings
at $M\sim 4$ TeV is very precise, more accurate even than in SusyGUTs.
This was shown, together with the robustness of the model, in \cite{Frampton:2002st}.\\
We believe grand unification at 4 TeV has no disadvantage relative to
unification at a trillion times higher scale, and has the advantage of avoiding
the dubious desert hypothesis.\\
To clarify the quiver theory construction, we explain in more detail the case
of the $Z_{12}$ orbifold by exhibiting the relevant quiver diagrams. In this
case the quiver diagram is a dodecahedron, like a clockface, with nodes
labeled as indicated in Eq. (\ref{quiver}).\\
Certain shortcuts make use of the symmetries of the quiver diagram and
obviate including every possible link which will make the diagram very
dense with links and more difficult to understand. Let us begin with the
chiral fermions which are denoted by oriented arrows between two
nodes. \\
The quarks can be counted by examining the $C \rightarrow W$
links and subtracting the $W \rightarrow C$ links, noting that anomaly freedom
dictates that the chiral fermions will necessarily appear only in the 
specific combination of Eq. (\ref{families}) and so no other $C$ links
need to be checked. The relevant quiver diagram is shown in Fig. 1. \\
\begin{figure}[h]
    \centering
\raisebox{-0.5\height}{\includegraphics[scale=0.3]{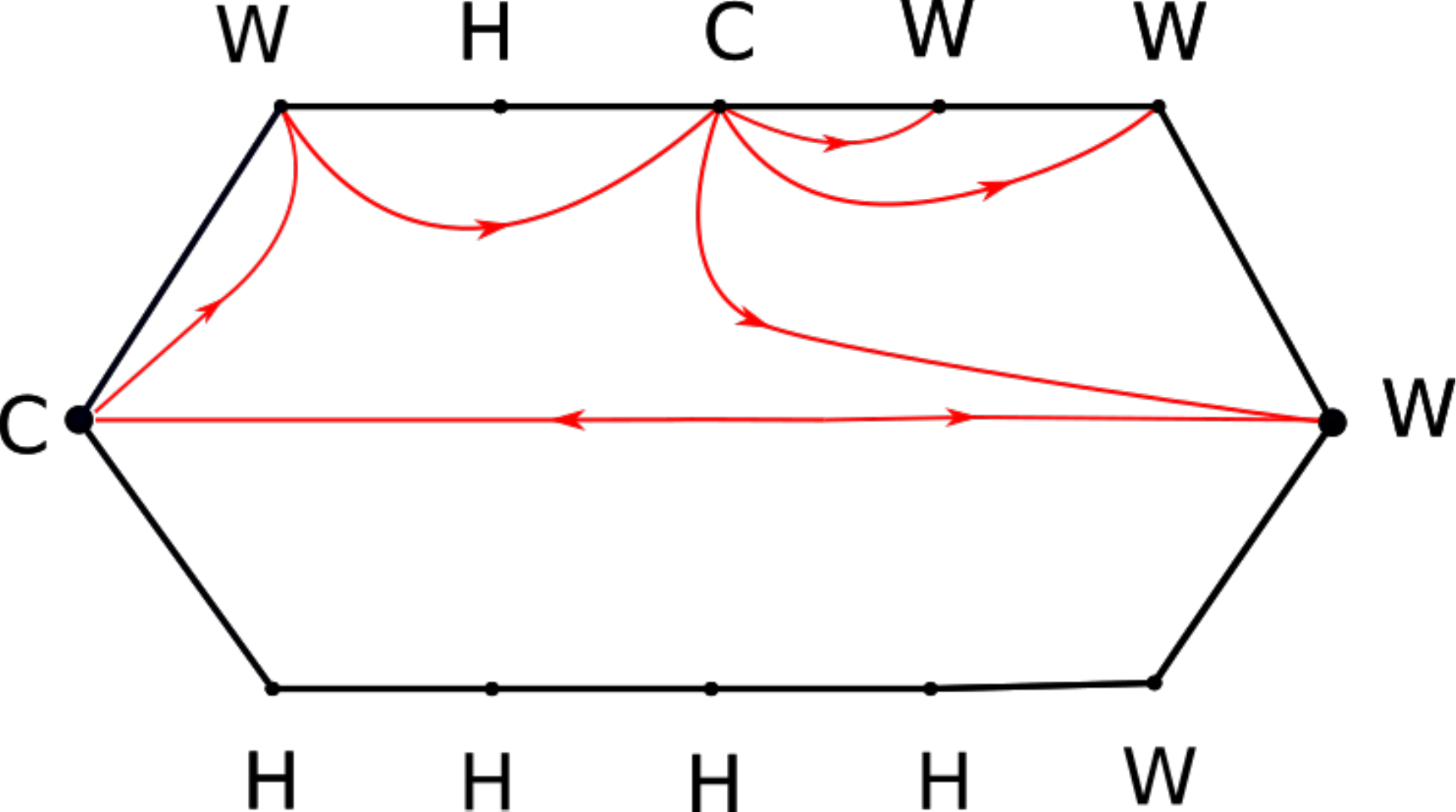}}
    \caption{ Quiver diagram showing quark states. }
    \label{oneq}
\end{figure}

\noindent
We see that there are five families and two antifamilies, resulting in
precisely three light chiral families as required. The family-antifamily
pairs are not chiral, but vector-like, and can therefore acquire Dirac
masses.\\
Next we exhibit in Fig. 2 and Fig. 3 two further $Z_{12}$ quiver diagrams
which illustrate the scalar sector. Complex scalars are denoted by unoriented
dashed lines. We must ensure and check that there are sufficient scalars
whose VEVs can spontaneously break the $SU(3)_C^2$ down to $SU(3)_C$,
the $SU(3)_W^5$ down to $SU(3)_W$ and finally $SU(3)_H^5$ down to
$SU(3)_H$.\\
\begin{figure}[h]
    \centering
\raisebox{-0.5\height}{\includegraphics[scale=0.3]{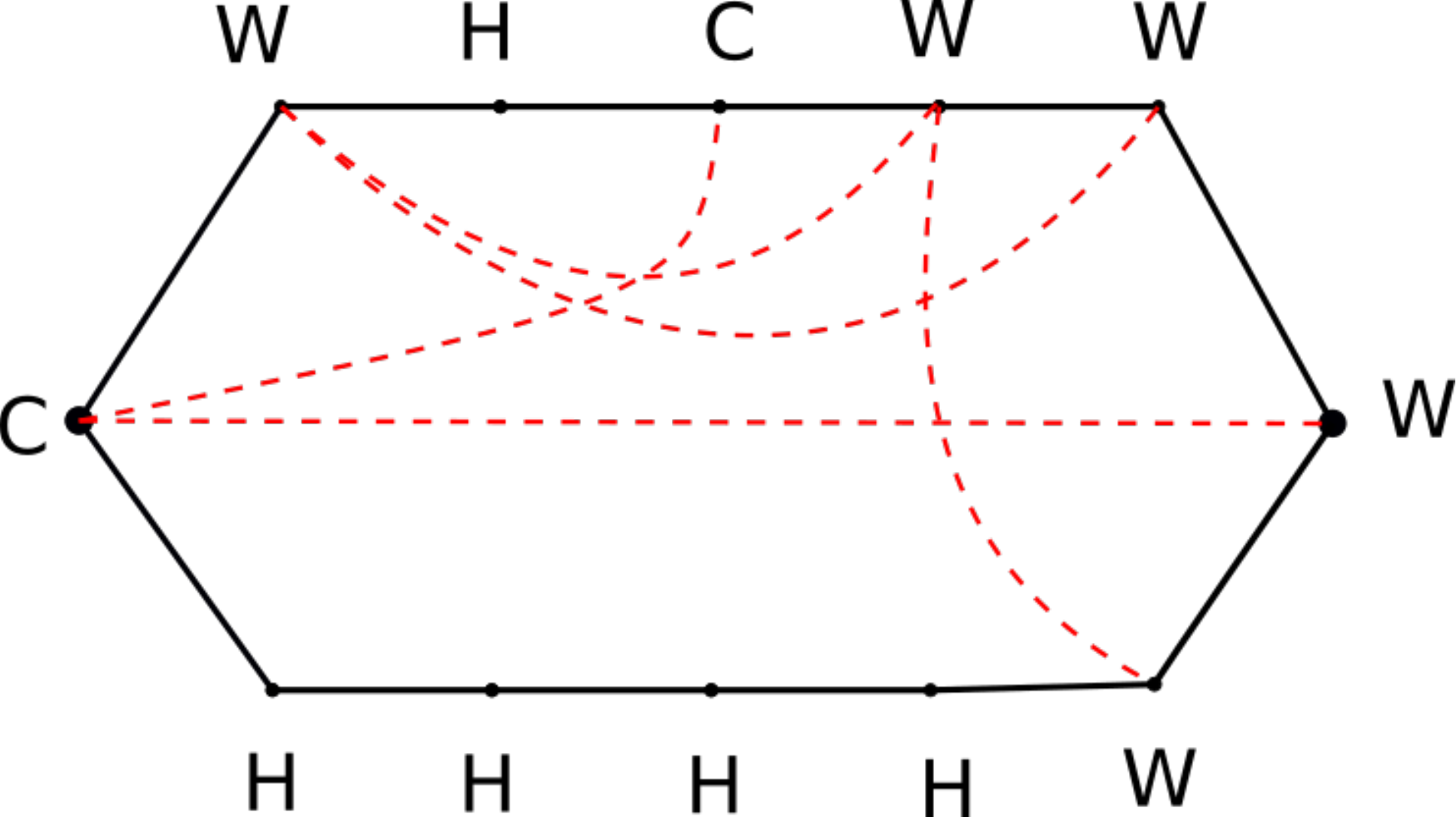}}
    \caption{Scalar states which break $SU(3)_W^5$}
    \label{twoq}
\end{figure}

\begin{figure}[h]
    \centering
\raisebox{-0.5\height}{\includegraphics[scale=0.3]{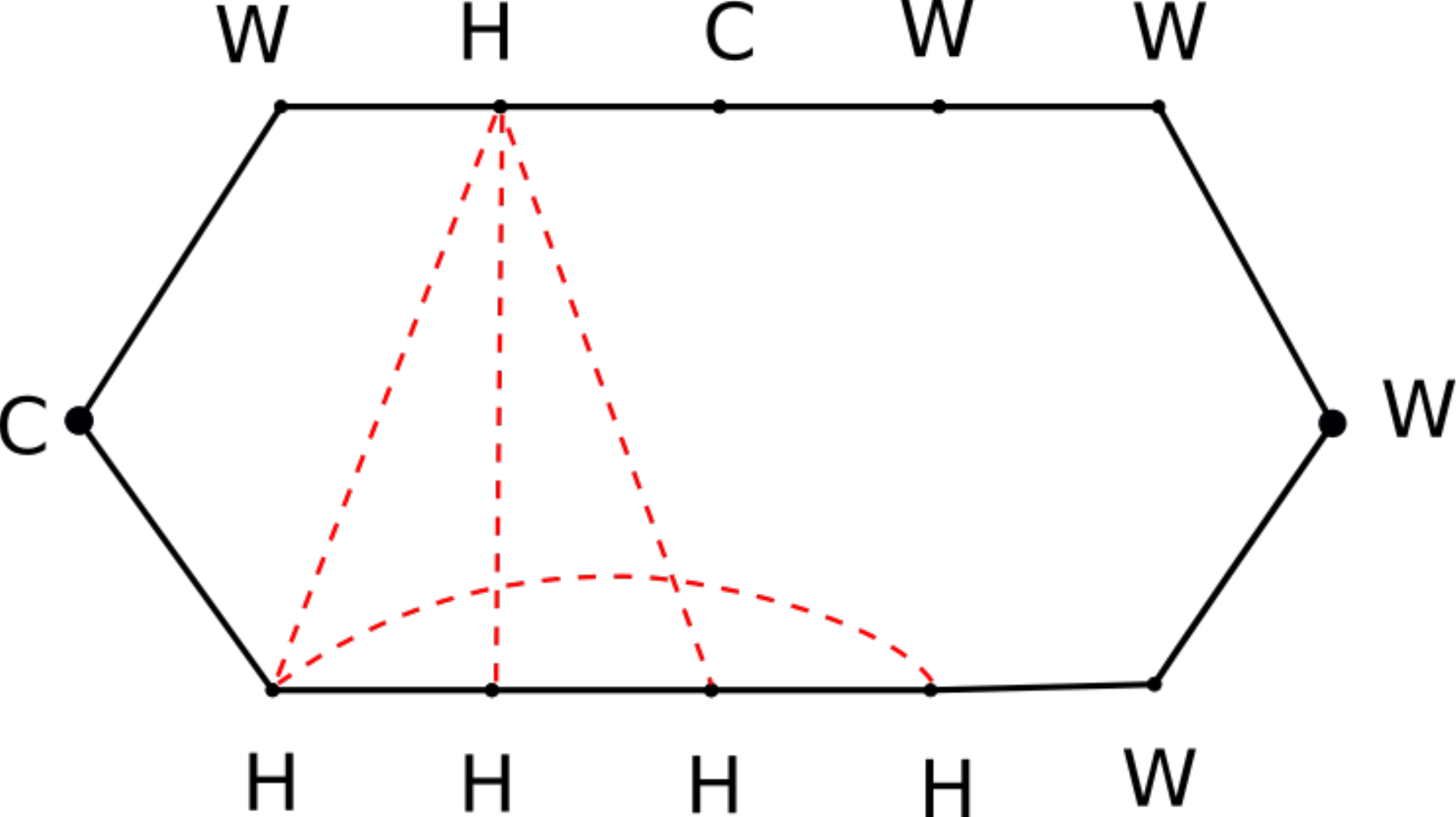}}
    \caption{scalar states that break $SU(3)_H^5$}
    \label{threeq}
\end{figure}

\noindent
To break any $SU(3)^n$ down to the diagonal subgroup, a necessary and
sufficient condition is that the bifundamental scalars link all $n$ of the
original gauge groups together. The $SU(3)$ gauge groups cannot be disconnected
into subgroups nor can the bifundamental scalars separate into disconnected
parts. In Fig. 2 the breakings of $SU(3)_C^2$ down to $SU(3)_C$ and of
$SU(3)_W^5$ down to $SU(3)_W$ are shown to satisfy all these tight
constraints so that the required spontaneous symmetry breaking is possible.\\
In Fig. 3 it is shown that the symmetry breaking $SU(3)_H^5$ down to
$SU(3)_H$ also satisfies the same unforgiving connectivity requirements. \\
We note that this symmetry breaking is very nontrivial and is what underlies
the correct identification of the nodes in Eq.(\ref{quiver}), which is unique in allowing
the required outcomes for both chiral fermions and complex scalars.\\
In fact, for $4$ TeV grand unification without a desert, the $SU(3)^{12}$
construction appears to be unique when one insists that one arrives at
three chiral families under trinification, and hence under the Standard Model,
as well as ensuring that correct symmetry breaking is permitted. \\
The GUT gauge group $SU(3)^{12}$ has dimension 96 which is
bigger than the dimensions 24 and 45 of $SU(5)$ and $SO(10)$
respectively. This can be regarded as the price to pay to avoid the
desert. The wealth of additional state at $4$TeV also changes
the nature of the phase transition $SU(3)^{12} \rightarrow SU(3)^3$
which can generate the gravitational waves studied in the next section.

\section{Gravitational Waves}

\noindent
In the presence of a cosmological FOPT a new phase begins to nucleate as the universe cools down, with the region inside the bubbles containing the new phase. The latent free energy released detonates the transition on the bubble wall. A scalar field 
acquires a vev in a state of true vacuum in the interior of the bubble, and of false vacuum in the exterior, causing the expansion and the collision of the bubble walls.  This takes place at a specific nucleation temperature $T_n$, which can be determined only by a combination of numerical simulations and an accurate study of the scalar potential, with the inclusion of thermal effects. The transition stops when the bubbles occupy all the volume.\\
Clearly, one expects a driving potential derived from the scalar sector which, in the simplest examples, is characterised by at least two separate scales. These scales identify the two local minima, separated by a maximum of considerable height, in order to guarantee a state of false vacuum of the system at the beginning of the nucleation phase \cite{Coleman:1977py,Callan:1977pt}\cite{Linde:1981zj}. The presence of a false vacuum with a sufficient amount of supercooling is a natural requirement for having a strongly FOPT.
In turn, these qualitative conditions point towards the possibility of having a significant emission of GWs. \\
As we have already mentioned, in the present quiver model, there is only one cosmological phase transition
at a scale above the electroweak scale, at an energy/temperature of 4 TeV. 
As we are going to show, the quiver model allows a certain conformal scalar potential which breaks to trinification $SU(3)^3$, and that can be identified in its symmetry structure starting from its symmetry content. Therefore, this breaking is expected to induce vacuum transitions, from a local metastable minimum which is sufficiently trapped, to the true vacuum. We are going to describe it below. 
An additional element which afftects the transition are thermal 
effects. As usual, they can be taken into account  by the inclusion of a Coleman-Weinberg term
  \cite{Coleman:1973jx} ${\cal V}_{Eff}^{CW}$ at one-loop level and a finite temperature contribution
  \cite{Dolan:1973qd} ${\cal V}_{Eff}^T$ to arrive at an effective
  potential
  \begin{equation}
  {\cal V}_{Eff} = {\cal V}_{Eff}^{Tree} + {\cal V}_{Eff}^{CW} + {\cal V}_{Eff}^T.
  \label{EffPot}
  \end{equation}
This defines the most general class of effective potentials discussed
\cite{Caprini:2015zlo,Weir:2017wfa,Caprini:2019egz}.  \\
In our case, if we denote by $\Phi$ (without labels) a generic scalar field taken from any of the
groups of scalar fields in Eq.(\ref{14scalars}), Eq.(\ref{25678scalars}) and
Eq.(\ref{39101112scalars}), the Coleman-Weinberg \cite{Coleman:1973jx} term in the potential ${\cal V}_{Eff}$ is then

\begin{equation}
{\cal V}_{Eff}^{CW} = \frac{\lambda}{4!} \Phi^4 + \frac{\lambda^2 \Phi^4}{256 \phi^2} \left( \ln \frac{\Phi^2}{M^2}  - \frac{25}{5} \right) 
\label{CW}
\end{equation}

\noindent
The Dolan-Jackiw-Weinberg finite temperature correction in terms of our generic scalar field $\Phi$
can be written \cite{Dolan:1973qd,Weinberg:1974hy} as

\begin{equation}
{\cal V}_{Eff}^T = \frac{\pi^2 T^4}{90}  + \frac{M^2 T^2}{24} - \frac{1}{12 \pi} M^3 T
-\frac{1}{64\pi^2}M^4 \ln M^2T^2 + \frac{c}{64 \pi^2} M^4 + O(M^6/T^2).
\label{DJW}
\end{equation} 
from which the effective potential is given by Eq. ({\ref{EffPot}). We will comment on the the structure of the potential at zero temperature below.  

\section{Specific features of the quiver theory}
Even in the absence of additional information about the way in which this transition takes place, due to the complexity of the scalar sector of the model, and within the assumption of a strongly FOPT, the goal of our analysis is to investigate the dependence of the GW emission on two peculiar parameters of the model, which are its large number of massless degrees of freedom and the relatively low transition temperature $T\sim 4 $ TeV.   
%Henceforth, without justification, we shall assume this to be strongly FOPT(First
%Order Phase Transition).

\noindent
We shall need $g_*$, the equivalent number of massless degrees degrees
of freedom for the quiver theory, defined by

\begin{equation}
g_* = n_B + \frac{7}{8} n_F
\label{g star}
\end{equation}
where $n_B, n_F$ is the number for bosons, fermions respectively.
It is easier to count $g_*$ before 
spontaneous symmetry breaking, although of course the result is the same.\\
In the standard model with three families we have

\begin{eqnarray}
n_B ({\rm spin} = 1) & = & 12 \times 2 = 24 \nonumber \\
n_B ({\rm spin} = 0) & = & 4 \nonumber \\
n_F ({\rm spin} = 1/2) & = & 3 \times 15 \times 2 = 90
\label{g star SM}
\end{eqnarray}
so that in this case
\begin{equation}
g_* = 28 + \frac{7}{8} (90) = 106.75
\label {g star SMexplicit}
\end{equation}
which will also be $g_*$ for the quiver theory at energies $E < 4$ TeV.\\
In our present $SU(3)^{12}$ quiver theory we recall from the previous section
that the scalars are in the bifundamental representations

\begin{equation}
\sum_{i=1}^{3}\sum_{\alpha=1}^{12} \left(3_{\alpha} ,\bar{3}_{\alpha+a_{i}} \right) 
\end{equation}
with $a_1 = 3,4,5$, and the chiral fermions are in bifundamentals

\begin{equation}
\sum_{\mu=1}^{4}\sum_{\alpha=1}^{12} \left(3_{\alpha}, \bar{3}_{\alpha+A_{\mu}} \right)
\end{equation}
with $A_{\mu}= 1,2,3,6$.

\noindent
The equivalent massless degrees of freedom are
\begin{eqnarray}
n_B({\rm spin}=1) & = & 96 \times 2 = 192 \nonumber \\
n_B({\rm spin}=0) & = & 12 \times 9 \times 3 = 324 \nonumber \\
n_F({\rm spin} = 1/2) & = & 12 \times 18 \times 4 = 864 \nonumber\\
\label{g star quiver}
\end{eqnarray}
so that for the full quiver theory
\begin{equation}
g_* = 516 + \frac{7}{8} (864) = 1,272
\label{G star quiver explicit}
\end{equation}
which is the number of effective massless degrees of freedom for $E \geq 4$ GeV.
In our present $SU(3)^{12}$ quiver theory we recall from the previous section
that the scalars are in the bifundamental representations

\begin{equation}
\sum_{i=1}^{3}\sum_{\alpha=1}^{12} \left(3_{\alpha} ,\bar{3}_{\alpha \pm a_{i}} \right) 
\label{Scalars}
\end{equation}
with $a_1 = 3,4,5$, and the chiral fermions are in bifundamentals

\begin{equation}
\sum_{\mu=1}^{4}\sum_{\alpha=1}^{12} \left(3_{\alpha}, \bar{3}_{\alpha+A_{\mu}} \right)
\end{equation}
with $A_{\mu}= 1,2,3,6$.\\
The equivalent massless degrees of freedom are
\begin{eqnarray}
n_B({\rm spin}=1) & = & 96 \times 2 = 192 \nonumber \\
n_B({\rm spin}=0) & = & 12 \times 9 \times 3 = 324 \nonumber \\
n_F({\rm spin} = 1/2) & = & 12 \times 18 \times 4 = 864 
\label{g star quiver}
\end{eqnarray}
so that for the full quiver theory
\begin{equation}
g_* = 516 + \frac{7}{8} (864) = 1,272
\label{G star quiver explicit}
\end{equation}
which is the number of effective massless degrees of freedom for $E \geq 4$ GeV.\\
We pause for few remarks. \\
%To discuss gravitational waves emitted during the phase transition at $T=4$ TeV
%in the early universe, we adopt some of the calculations presented in
%\cite{Caprini} and focus initially on the gravitational radiation from
%bubble collisions, assuming that we are dealing with a FOPT (First Order
%Phase Transition).\\
The nature of the phase transition depends on the effective potential
of the theory. Eq.(\ref{Scalars}) exhibits the scalars present in the quiver
and the twelve nodes of the quiver are identified in Eq.(\ref{quiver}). The
dodecahedral quiver has nodes which we label clockwise by 1 to 12
by Color(C), Weak (W) and Hypercharge (H) as follows:

\begin{equation}
(1)\, C\to  (2)\, W\to (3)\, H \to (4)\, C\to (5 - 8)\, W\to (9-12) \, H
\label{Nodes}
\end{equation}

\noindent 
We are initially concerned with the breaking $SU(3)^{12} \rightarrow SU(3)^3$
at scale $E = 4$ TeV. This can be studied separately for C, W and H in Eq.(\ref{Nodes}).\\
Let us define lower-case Greek indices $\alpha_i, \beta_i, \gamma_i, \delta_i \ldots
= 1,2,3$ for the SU(3) group of the $i^{th}$ node and discriminate between
subscripts which represent defining representations and superscripts
which denote anti-defining representations.\\
From Eq. (\ref{Nodes}) the SM color gauge group arises from the diagonal
subgroup of the SU(3)'s at nodes 1 and 4 respectively, and this symmetry
breaking is achieved by VEVs of the complex scalar bifundamentals:

\begin{equation}
\Phi_{\alpha_1}^{\beta_4} ~~ {\rm and} ~~  \Phi_{\alpha_4}^{\beta_1}
\label{14scalars}
\end{equation}
In the effective potential at tree level there are quadratic and quartic
terms involving the 1 to 4 bifundamentals as follows

\begin{eqnarray}
{\cal V}_{Eff}^{(C, Tree)} & = & {\cal C}^{(14)}_2 \left( \Phi_{\alpha_1}^{\beta_4}\Phi_{\alpha_1}^{\beta_4} \right) 
  + {\cal C}^{(14)}_4 \left( \Phi_{\alpha_1}^{\beta_4} \Phi_{\alpha_1}^{\beta_4} \right)^2 
  +  {\cal C}^{(14) '}_4 \left( \Phi_{\alpha_1}^{\beta_4} \Phi_{\beta_4}^{\gamma_1}
  \Phi_{\gamma_1}^{\delta_4}\Phi_{\delta_4}^{\alpha_1}  \right)
\label{V14}
\end{eqnarray}
To break to the trinification group
\begin{equation}
SU(3)_C \times SU(3)_W \times SU(3)_H
\label{trinification}
\end{equation}
a similar combination of bifundamental scalars conspire to arrive at diagonal subgroups for
both the five $SU(3)_W$ nodes
and the five $SU(3)_H$ nodes respectively. Another intermediate symmetry-breaking stage
is where $SU(3)_W$ in Eq.(\ref{trinification}) breaks to the $SU(2)_L$ of the SM, also
$SU(3)_W \times SU(3)_H$ breaks to the $U(1)_Y$ of the SM but for our analysis of
gravitational radiation we shall focus only on a FOPT where the quiver group $SU(3)^{12}$
breaks at $E=4$ TeV to the trinification group in Eq.(\ref{trinification}).\\
For W we use scalars connecting nodes 2-5-6-7-8 and the relevant scalar bifundamentals in Eq.(\ref{Scalars})
are
\begin{equation}
\Phi_{\alpha_2}^{\beta_5} ~~,~~ \Phi_{\alpha_2}^{\beta_6} ~~, ~~ \Phi_{\alpha_2}^{\beta_7} ~~ {\rm and} ~~ \Phi_{\alpha_5}^{\beta_8}
\label{25678scalars}
\end{equation}
The corresponding quadratic and quartic terms in the tree-level effective potential composed
of the scalars in Eq.(\ref{25678scalars}) are

\begin{eqnarray}
{\cal V}_{Eff}^{(W,Tree)} & = & {\cal C}^{(25678)}_2 \left( \Phi_{\alpha_2}^{\beta_5} \Phi_{\beta_5}^{\alpha_2} + \Phi_{\alpha_2}^{\beta_6} \Phi_{\beta_6}^{\alpha_2} +
\Phi_{\alpha_2}^{\beta_7} \Phi_{\beta_7}^{\alpha_2} +
\Phi_{\alpha_5}^{\beta_8} \Phi_{\beta_8}^{\alpha_5} \right)  \nonumber \\
&&
 +  {\cal C}^{(25678)}_4 \left( \Phi_{\alpha_2}^{\beta_5} \Phi_{\beta_5}^{\alpha_2} + \Phi_{\alpha_2}^{\beta_6} \Phi_{\beta_6}^{\alpha_2} +
\Phi_{\alpha_2}^{\beta_7} \Phi_{\beta_7}^{\alpha_2}  +\Phi_{\alpha_5}^{\beta_8} \Phi_{\beta_8}^{\alpha_5} \right)^2 \nonumber \\
&& + {\cal C}^{(25678) ' }_4 \left( \Phi_{\alpha_2}^{\beta_5} \Phi_{\beta_5}^{\gamma_2}
  \Phi_{\gamma_2}^{\delta_5}\Phi_{\delta_5}^{\alpha_2}  +
   \Phi_{\alpha_2}^{\beta_6} \Phi_{\beta_6}^{\gamma_2}
  \Phi_{\gamma_2}^{\delta_6}\Phi_{\delta_6}^{\alpha_2} \right.  
 \left. +  \Phi_{\alpha_2}^{\beta_7} \Phi_{\beta_7}^{\gamma_2}
  \Phi_{\gamma_2}^{\delta_7 }\Phi_{\delta_7}^{\alpha_2} +
   \Phi_{\alpha_5}^{\beta_8} \Phi_{\beta_8}^{\gamma_5}
  \Phi_{\gamma_5}^{\delta_8}\Phi_{\delta_8}^{\alpha_5} \right) \nonumber \\
\end{eqnarray}
For H we use scalars connecting nodes 3-9-10-11-12 and the relevant scalar bifundamentals in Eq.(\ref{Scalars})
are.
\begin{equation}
\Phi_{\alpha_3}^{\beta_{10}} ~~,~~ \Phi_{\alpha_3}^{\beta_{11}} ~~, ~~ \Phi_{\alpha_3}^{\beta_{12}} ~~ {\rm and} ~~ \Phi_{\alpha_9}^{\beta_{12}}
\label{39101112scalars}
\end{equation}

\noindent
The corresponding quadratic and quartic terms in the tree-level effective potential composed
of the scalars in Eq.(\ref{39101112scalars}) are

\begin{eqnarray}
{\cal V}_{Eff}^{(H,Tree)} & = & {\cal C}^{(39101112)}_2 \left( \Phi_{\alpha_3}^{\beta_{10}} \Phi_{\beta_{10}}^{\alpha_3} + \Phi_{\alpha_3}^{\beta_{11}} \Phi_{\beta_{11}}^{\alpha_3} +
\Phi_{\alpha_3}^{\beta_{12}} \Phi_{\beta_{12}}^{\alpha_3} +
\Phi_{\alpha_9}^{\beta_{12}} \Phi_{\beta_{12}}^{\alpha_9} \right)  \nonumber \\
&& +  {\cal C}^{(39101112)}_4 \left(  \Phi_{\alpha_3}^{\beta_{10}} \Phi_{\beta_{10}}^{\alpha_3} + \Phi_{\alpha_3}^{\beta_{11}} \Phi_{\beta_{11}}^{\alpha_3} +
\Phi_{\alpha_3}^{\beta_{12}} \Phi_{\beta_{12}}^{\alpha_3} +
\Phi_{\alpha_9}^{\beta_{12}} \Phi_{\beta_{12}}^{\alpha_9} \right)^2  \nonumber \\
& & + {\cal C}^{(39101112) ' }_4 \left( \Phi_{\alpha_3}^{\beta_{10}} \Phi_{\beta_{10}}^{\gamma_3}
  \Phi_{\gamma_3}^{\delta_{10}}\Phi_{\delta_{10}}^{\alpha_3}  +
   \Phi_{\alpha_3}^{\beta_{11}} \Phi_{\beta_{11}}^{\gamma_3}
  \Phi_{\gamma_3}^{\delta_{11}}\Phi_{\delta_{11}}^{\alpha_3} \right.  \nonumber \\
  & & ~~~~~~~~~~~~~~ \left. +  \Phi_{\alpha_3}^{\beta_{12}} \Phi_{\beta_{12}}^{\gamma_3}
  \Phi_{\gamma_3}^{\delta_{12} }\Phi_{\delta_{12}}^{\alpha_3} +
   \Phi_{\alpha_9}^{\beta_{12}} \Phi_{\beta_{12}}^{\gamma_9}
  \Phi_{\gamma_9}^{\delta_{12}}\Phi_{\delta_{12}}^{\alpha_9} \right) \nonumber \\
\end{eqnarray}

\noindent
Because the quiver theory above 4 TeV is conformal we must impose
$C_2 = 0$ in all the quadratic terms. Next, before adding the three 
${\cal V}_{Eff}^{Tree}$ expressions, let us examine
the symmetries of the dodecahedral quiver which imply that

\begin{eqnarray}
C_4^{(25678)} & = & C_4^{(39101112)} \equiv D_4 \nonumber \\
C_4^{(25678)'} & = & C_4^{(39101112)'}\equiv D_4^{'} 
\end{eqnarray}
whereupon, suppressing superscripts, the most general tree-level
effective potential is

\begin{eqnarray}
{\cal V}_{Eff}^{Tree}  =  {\cal C}_4 \left( \Phi_{\alpha_1}^{\beta_4} \Phi_{\alpha_1}^{\beta_4} \right)^2  +  {\cal C}^{'}_4 \left( \Phi_{\alpha_1}^{\beta_4} \Phi_{\beta_4}^{\gamma_1}
 \Phi_{\gamma_1}^{\delta_4}\Phi_{\delta_4}^{\alpha_1}  \right)   
   + {\cal D}_4 \left( \Phi_{\alpha_2}^{\beta_5} \Phi_{\beta_5}^{\alpha_2} + \Phi_{\alpha_2}^{\beta_6} \Phi_{\beta_6}^{\alpha_2} +
\Phi_{\alpha_2}^{\beta_7} \Phi_{\beta_7}^{\alpha_2} +
\Phi_{\alpha_5}^{\beta_8} \Phi_{\beta_8}^{\alpha_5} \right)^2  && \nonumber \\
 + {\cal D}^{' }_4 \left( \Phi_{\alpha_2}^{\beta_5} \Phi_{\beta_5}^{\gamma_2}
  \Phi_{\gamma_2}^{\delta_5}\Phi_{\delta_5}^{\alpha_2}  +
   \Phi_{\alpha_2}^{\beta_6} \Phi_{\beta_6}^{\gamma_2}
  \Phi_{\gamma_2}^{\delta_6}\Phi_{\delta_6}^{\alpha_2} 
 +  \Phi_{\alpha_2}^{\beta_7} \Phi_{\beta_7}^{\gamma_2}
  \Phi_{\gamma_2}^{\delta_7 }\Phi_{\delta_7}^{\alpha_2} +
   \Phi_{\alpha_5}^{\beta_8} \Phi_{\beta_8}^{\gamma_5}
  \Phi_{\gamma_5}^{\delta_8}\Phi_{\delta_8}^{\alpha_5} \right). && \nonumber \\
  \end{eqnarray}
  
%\section{GWs and PTs with a conformal quiver model}

\section{Production of GW and the parameters choice}
\noindent
As mentioned in the previous sections, the two main features of our quiver model 
are the large number of massless degrees of freedom present at the phase transition and the relatively low temperature at which the unification of the gauge couplings is reached. Given the complexity of the scalar sector of the model, it is beyond the scope of the current analysis to provide further details about the way the conformal symmetry is broken, with the generation of appropriate scales in the potential which would allow vacuum and thermal transitions of significant strength. As already mentioend, we will simply assume that this is possible, leaving a discussion of this issue to future work. \\
We recall that the energy density of the gravitational wave is measured (today) by the variables
\begin{equation}
h_0^2 \Omega_{GW}(f)\equiv\left( \frac{h_0^2}{\rho_c}\frac{d\,\rho_{GW}}{d\,\log\, f}\right)_0
\label{eqx}
\end{equation}
expressed in frequency $(f)$ octaves, in which we are going to separate the various contributions.\\
Thus we may write, for the final contribution to the energy density, as a fraction of the critical
density:

\begin{equation}
\Omega_{GW} (f) = \Omega^{Coll}_{GW} (f) + \Omega^{SW}_{GW} (f) + \Omega^{Turb}_{GW} (f)
\label{sources}
\end{equation}
where the terms on the right hand side denote contributions sourced by bubble collisions, sound waves, and plasma 
turbulence, respectively.

\bigskip

\noindent
For the contribution $\Omega_{GM}^{Coll}(f)$ from the bubble collisions, we may
write, when $\beta/H^* \gg 1$,

\begin{equation}
\Omega_{GW}^{Coll} (f) = \Omega_{GW}^{Coll} (f_{peak}) S^{Coll} (f)
\end{equation}
where the spectral function is given by \cite{Huber:2008hg}
\begin{equation}
S^{Coll}(f) = \frac{(a+b) f_{Peak}f^a} {bf_{Peak}^{a+b} + af^{a+b}}
\end{equation}
where $(a,b) \simeq (3,1.0)$. The peak amplitude is provided by \cite{Kosowsky:1992vn}
\begin{equation}
h^2 \Omega_{GW}^{Coll} (f_{Peak}) \simeq 1.7 \times 10^{-5} \kappa^2 \Delta 
\left( \frac{\beta}{H_*} \right)^{-2} \left(\frac{\alpha}{1+\alpha} \right)^2 
\left( \frac{g_*}{100} \right)^{-\frac{1}{3}}
\label{amplitude}
\end{equation}
where the efficiency factor $\kappa$ was first derived by Steinhardt \cite{Steinhardt:1981ct} as
\begin{equation}
\kappa= \frac{1}{1+A\alpha} \left( A\alpha + \frac{4}{27} \sqrt{\frac{3 \alpha}{2}} \right)
\end{equation}
with $A=0.715$.

\bigskip

\noindent
The peak frequency in Eq.(\ref{amplitude}) is given by
\begin{equation}
f_{Peak} \simeq 17 \left(\frac{f_*}{\beta}\right) \left(\frac{\beta}{H_*} \right)
\left(\frac{T_*}{10^8 GeV} \right)\left(
\frac{g_*}{100} \right)^{\frac{1}{6}} {\rm Hz}
\end{equation}
\begin{equation}
\frac{f_*}{\beta} = \frac{0.62}{1.8 -0.1v_b + v_b^2}
\end{equation}
while in the same equation the dependence of $\Delta$ on the velocity
$v_b$ of the bubble wall is given by \cite{Steinhardt:1981ct,Espinosa:2010hh,Espinosa:2011ax,Dorsch:2018pat}

\begin{equation}
v_b(\alpha)=\frac{\frac{1}{\sqrt{3}}+\sqrt{\alpha^2+\frac{2\alpha}{3}}}{1+\alpha},
\end{equation}
with

\begin{equation}
\Delta = \frac{0.11v_b^3}{0.42 + v_b^2}.
\end{equation}

\bigskip

\noindent
For the second term in Eq.(\ref{sources}) we may similarly write

\begin{equation}
\Omega_{GW}^{SW} (f) = \Omega_{GW}^{SW} (f_{peak}) S^{SW} (f)
\label{SW}
\end{equation}
with \cite{Hindmarsh:2013xza, Hindmarsh:2015qta, Espinosa:2010hh}

\begin{equation}
h^2 \Omega_{GW}^{SW} (f_{Peak}) \simeq 2.7 \times 10^{-6} \kappa_v^2 v_b
\left(
\frac{\beta}{H_*} \right)^{-1} \left(\frac{\alpha}{1+\alpha} \right)
\left(\frac{g_*}{100} \right)^{-\frac{1}{3}} (H_*\tau_{SW})
\label{amplitudeSW}
\end{equation}
\begin{equation}
\kappa_v \simeq \frac{\alpha}{0.73+0.083\sqrt{\alpha} +\alpha}.
\label{kappaSW}
\end{equation}

\bigskip

\noindent
According to \cite{Huber:2008hg} the peak sound wave frequency is provided by
\begin{equation}
f_{Peak} \simeq 19 \frac{1}{v_b} \left(\frac{\beta}{H_*} \right) \left(\frac{T_*}{10^8 GeV} \right)
\left(\frac{g_*}{100} \right)^{\frac{1}{6}}    {\rm Hz}
\label{PeakSW}
\end{equation}
while, according to \cite{Caprini:2015zlo}, the sound-wave spectral function 
in Eq.(\ref{SW}) is
\begin{equation}
S^{SW}(f) =
\left(\frac{f}{f_{Peak}} \right)^3 
\left( \frac{7}{4+3\left(\frac{f}{f_{Peak}} \right)^2} \right)^{\frac{7}{2}}.
\label{spectralSW}
\end{equation}

\bigskip

\noindent
The sound waves remain active for a time $\tau_{SW}$
\begin{equation}
\tau_{SW} = \frac{R_*}{U_f}
\end{equation}
in which $R_*$ is the root mean bubble separation
$R_* \simeq (8\pi)^{\frac{1}{3}} v_b/\beta$
and $U_f$ is the root mean square of the fluid velocity \cite{Hindmarsh:2015qta}
\begin{equation}
U_f^2 \simeq \frac{3}{4} \left(\frac{\alpha}{1+\alpha} \right) \kappa_v.
\end{equation}

\noindent
We note that the last factor in Eq. (\ref{amplitudeSW}) represents an important comparison
of the sonic period to the Hubble time of cosmic expansion\cite{Ellis:2018mja,Ellis:2019oqb,Cutting:2019zws,Hindmarsh:2019phv,Ellis:2020awk,Schmitz:2020rag}.

\bigskip

\noindent
For the third and final term in Eq.(\ref{sources}) representing plasma turbulence, we write
similarly again:

\begin{equation}
\Omega_{GW}^{Turb} (f) = \Omega_{GW}^{Turb} (f_{peak}) S^{Turb} (f)
\label{Turb}
\end{equation}
in which the factors are given by the estimates \cite{Kamionkowski:1993fg}
\begin{equation}
h^2 \Omega_{GW}^{Turb} (f_{Peak}) \simeq 3.4\times10^{-4} v_b \left(\frac{\beta}{H_*} \right)^{-1}
\left(\frac{\kappa_{Turb} \alpha}{1+\alpha} \right)^{\frac{3}{2}} \left(\frac{g_*}{100} \right)^{-\frac{1}{3}}
\end{equation}
\begin{equation}
f_{Peak} \simeq 37 \frac{1}{v_b} \left( \frac{\beta}{H_*} \right) \left(\frac{T_*}{10^8 GeV} \right)
\left( \frac{g_*}{100} \right)^{\frac{1}{6}}.  {\rm Hz}.
\end{equation}

\noindent
The spectral function $S_{GW}^{Turb}(f)$ in Eq.(\ref{Turb}) is provided by
\cite{Caprini:2015zlo,Caprini:2009yp,Binetruy:2012ze}
\begin{equation}
S^{Turb}(f) = \frac { \left( \frac{f}{f_{Peak}} \right)^3 } 
{\left(1 + \frac{f}{f_{Peak}} \right)^{11/3} 
\left(1+\frac{8\pi f}{h_*} \right)}
\end{equation}
\begin{equation}
h_* = 17 \left(\frac{T_*}{10^8 GeV} \right) \left(\frac{g_*}{100} \right)^{\frac{1}{6}} {\rm Hz.}
\end{equation}
wherein we set $\kappa_{Turb} \simeq 0.05 \kappa_v$.\\
In order to derive numerical predictions for the peak frequency of the quiver model and compare the results with other models, we pause for some considerations. One of the most important parameters appearing in all the equations presented above is 
$\beta/H_*$, which is derived from the tunneling action around the time when the transition occurs $(t_*)$ at the temperature $T_*$, using the adiabatic time-temperature relation
\begin{equation}
\frac{d t}{dT}=-\frac{1}{T H(T)},
\end{equation}
in the form \cite{Kosowsky:1991ua,Kosowsky:1992vn}\cite{Turner:1992tz}
\begin{equation} 
S(t)=S(t_*) -\beta (t- t_*) +  O((t -t _*)^2)    \qquad \qquad \textrm{with}
\qquad \frac{\beta}{H_*}= T_* \frac{d S}{dT}\vline_{T=T_*}. 
\end{equation}
$\beta=\dot{\Gamma}/\Gamma_*$ measures the time variation of the nucleation rate and $\tau\equiv 1/\beta$ characterizes the time scale of the phase transition at the transition time $t_*$. The parameter $\beta/H_*$ is defined by the ratio between $\tau$ and the Hubble time $1/H_*$. It is one of the most important parameters controlling the energy released into GWs at the phase transition. A more in depth analysis shows that one should set a distinction between the thermal and the vacuum tunneling contributions to $\Gamma(t)$, which involves either the three dimensional $S_3$ or four dimensional $S_4$ Euclidean bouncing solutions, which will not be of our concern in this work, as well as a finer characterization of the nucleation temperature (see for instance \cite{Ellis:2020awk}). \\ 
If we denote the nucleation rate with $\Gamma(t)$, 
the temperature of the transition is defined to be the temperature at which the probablility of nucleating one bubble per Hubble volume per Hubble time is one
\begin{equation}
\Gamma(t)\sim T^4 e^{-S(t)}\qquad \qquad \frac{\Gamma}{H^4}\sim 1 
\end{equation} 
which gives for the tunneling action the expression  
\begin{equation}
S(T_*)\sim - 4\log \frac{T_*}{m_P}, 
\end{equation}
where $M_{P}$ is the Planck mass. As an order of magnitude estimate one can set $\beta\sim H_* S_*$ \cite{Kamionkowski:1993fg,Hogan:1986qda} which gives a value $\beta/H_*\sim O(10^2)$ at the electroweak scale, and is a good approximation also in our case, due to the logarithmic dependence of $S(T_*)$ on $T_*$.\\  
Therefore we set
\begin{equation}
 \frac{\beta}{H_*}\sim 100-300,\qquad \qquad T_*\sim 4000\, \textrm{GeV},\qquad \qquad g_*=1732
 \end{equation}
and vary $\alpha$, the strength of the PT.
\section{Results}
\begin{table}[ht]
\caption{Numerical values of the peak frequency and GW emissions  for the collisional contributions. } % title of Table
\centering % used for centering table
\begin{tabular}{c c c c} % centered columns (6 columns)
\hline\hline %inserts double horizontal lines
$\beta/H_*$ & $\alpha$  &  $f_{peak}(Hz)$ &$h_0^2\Omega_{coll}$ \\ [0.5ex] % inserts table
%heading
\hline % inserts single horizontal line
100 & 0.60   & 2.7$\times 10^{-2}$   &$0.97\times 10^{-12}$ \\
100 & 0.65  & 2.7$\times 10^{-2}$   &$1.2\times 10^{-12}$ \\
100 & 0.70   & 2.7$\times 10^{-2}$   & $1.4\times 10^{-12}$ \\
100 & 0.75  & 2.7$\times 10^{-2}$    & $1.7\times 10^{-12}$ \\
100 & 0.80   & 2.6$\times 10^{-2}$    & $1.9\times 10^{-12}$ \\ 

200 & 0.60   & 5.4$\times 10^{-2}$     & $2.41\times 10^{-13}$ \\
200 & 0.65  & 5.3$\times 10^{-2}$   &$2.9\times 10^{-13}$ \\
200 & 0.70   & 5.3$\times 10^{-2}$     & $3.5\times 10^{-13}$ \\
200 & 0.75  & 5.3$\times 10^{-2}$    & $4.1\times 10^{-13}$ \\
200 & 0.80   & 5.3$\times 10^{-2}$     & $4.8\times 10^{-13}$ \\ 

300 & 0.60   & 8.0$\times 10^{-2}$     & $1.1\times 10^{-13}$ \\
300 & 0.65  & 8.0$\times 10^{-2}$     & $1.3\times 10^{-13}$ \\
300 & 0.70   & 8.0$\times 10^{-2}$      & $1.6\times 10^{-13}$ \\ 
300 & 0.75  & 8.0$\times 10^{-2}$      & $1.8\times 10^{-13}$ \\
300 & 0.80   & 8.9$\times 10^{-2}$     & $2.1\times 10^{-13}$ \\[1ex] % [1ex] adds vertical space
\hline %inserts single line
\end{tabular}
 % is used to refer this table in the text
\end{table}

\begin{table}[ht]
\caption{Numerical values for the PT parameters for the sound waves contributions. } % title of Table
\centering % used for centering table
\begin{tabular}{c c c c c c c} % centered columns (6 columns)
\hline\hline %inserts double horizontal lines
$\beta/H_*$ & $\alpha$ &   $f_{peak}(Hz)$  &$h_0^2\Omega_{sw}$ \\ [0.5ex] % inserts table
%heading
\hline % inserts single horizontal line
100 & 0.60   & $1.3\times 10^{-1}$     &$1.9\times 10^{-11}$ \\
100 & 0.65  & $1.3\times 10^{-1}$   &$2.2\times 10^{-11}$ \\
100 & 0.70   & $1.4\times 10^{-1}$     &$2.5\times 10^{-11}$ \\
100 & 0.75  & $1.3\times 10^{-1}$      & $2.8\times 10^{-11}$ \\
100 & 0.80   & $1.3\times 10^{-1}$     & $3.1\times 10^{-11}$ \\ 

200 & 0.60   & $2.7 \times 10^{-1}$    & $4.7\times 10^{-12}$ \\
200 & 0.65  & $2.7\times 10^{-1}$   & $5.5\times 10^{-12}$ \\
200 & 0.70   & $2.7\times 10^{-1}$     & $6.2\times 10^{-12}$ \\
200 & 0.75  & $2.7 \times 10^{-1}$    & $7.0\times 10^{-12}$ \\
200 & 0.80   & $2.6 \times 10^{-1}$    & $7.8\times 10^{-12}$ \\ 

300 & 0.60   & $4.0\times 10^{-1}$     & $2.1\times 10^{-12}$ \\
300 & 0.65  & $4.0\times 10^{-1}$     & $2.4\times 10^{-12}$ \\
300 & 0.70   & $4.0\times 10^{-1}$    & $2.8\times 10^{-12}$ \\ 
300 & 0.75  & $4.0\times 10^{-1}$     & $3.1\times 10^{-12}$ \\
300 & 0.80   & $4.0\times 10^{-1}$     & $3.5\times 10^{-12}$ \\[1ex] % [1ex] adds vertical space
\hline %inserts single line
\end{tabular}
 % is used to refer this table in the text
\end{table}

\begin{table}[ht]
\caption{Numerical values for the PT parameters for the turbulence contributions. } % title of Table
\centering % used for centering table
\begin{tabular}{c c c c c c c} % centered columns (6 columns)
\hline\hline %inserts double horizontal lines
$\beta/H_*$ & $\alpha$ &   $f_{peak}(Hz)$  &$h_0^2\Omega_{turb}$ \\ [0.5ex] % inserts table
%heading
\hline % inserts single horizontal line
100 & 0.60   &  $1.9\times 10^{-1}$   &$8.7\times 10^{-10}$ \\
100 & 0.65  &   $1.9\times 10^{-1}$       &$1.0\times 10^{-10}$ \\
100 & 0.70   & $1.9\times 10^{-1}$     & $1.1\times 10^{-9}$ \\
100 & 0.75  & $1.9\times 10^{-1}$    & $1.3\times 10^{-9}$ \\
100 & 0.80   & $1.9\times 10^{-1}$    & $1.4\times 10^{-9}$ \\ 

200 & 0.60   & $3.8\times 10^{-1}$    & $4.3\times 10^{-10}$ \\
200 & 0.65  & $3.8 \times 10^{-1}$   &$5.0\times 10^{-10}$ \\
200 & 0.70   & $3.8\times 10^{-1}$    & $5.7\times 10^{-10}$ \\
200 & 0.75  & $3.8\times 10^{-1}$    & $6.4\times 10^{-10}$ \\
200 & 0.80   & $3.8\times 10^{-1}$    & $7.1\times 10^{-10}$ \\ 

300 & 0.60   & $5.8\times 10^{-1}$     & $2.9\times 10^{-10}$ \\
300 & 0.65  & $5.8\times 10^{-1}$      & $3.3\times 10^{-10}$ \\
300 & 0.70   & $5.7 \times 10^{-1}$     & $3.8\times 10^{-10}$ \\ 
300 & 0.75  & $5.7\times 10^{-1}$     & $4.2\times 10^{-10}$ \\
300 & 0.80   & $5.7\times 10^{-1}$    & $4.7\times 10^{-10}$ \\[1ex] % [1ex] adds vertical space
\hline %inserts single line
\end{tabular}
 % is used to refer this table in the text
\end{table}

\begin{figure}[ht]
    \centering
\raisebox{-0.5\height}{\includegraphics[scale=0.4]{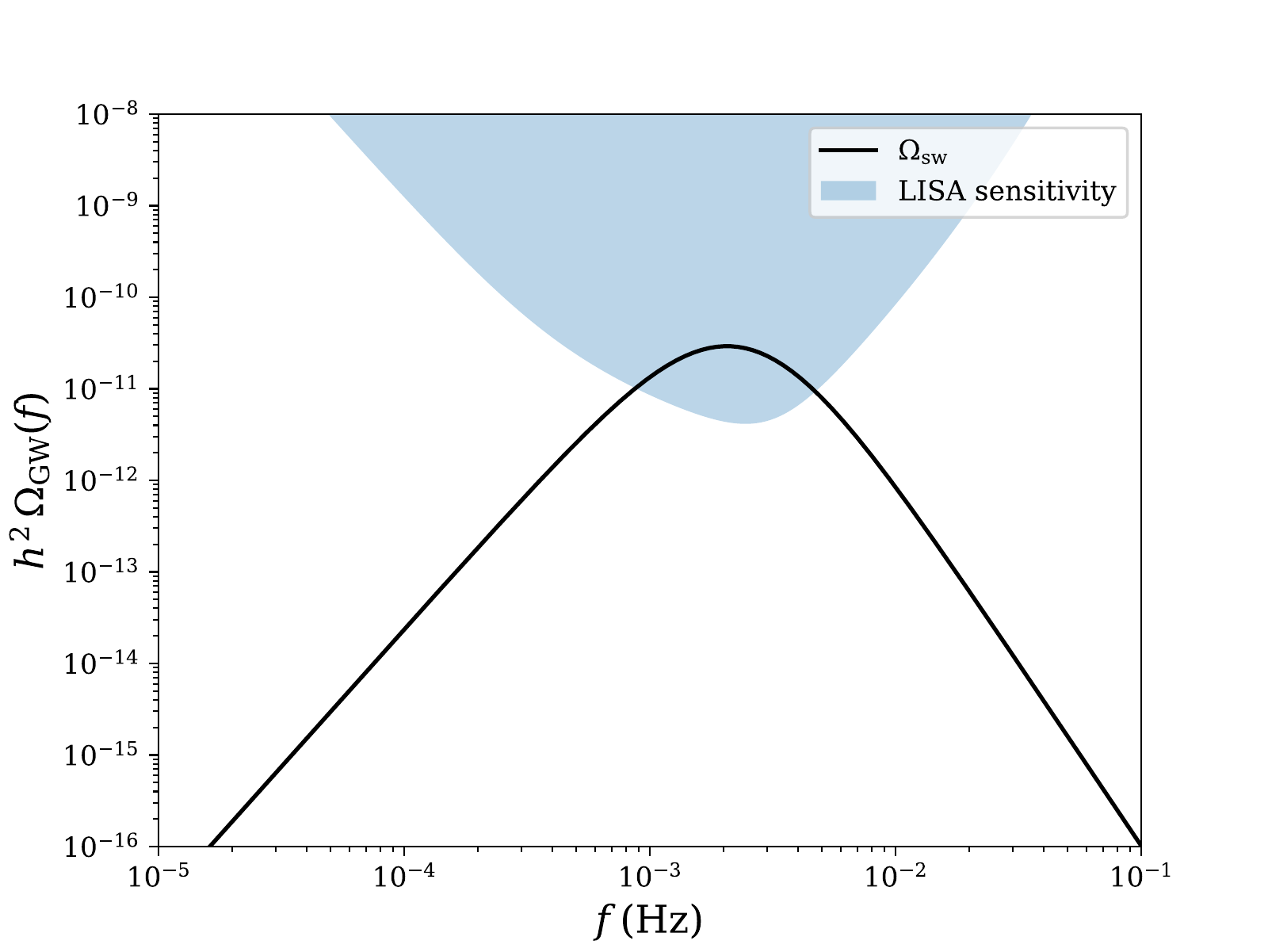}\includegraphics[scale=0.4]{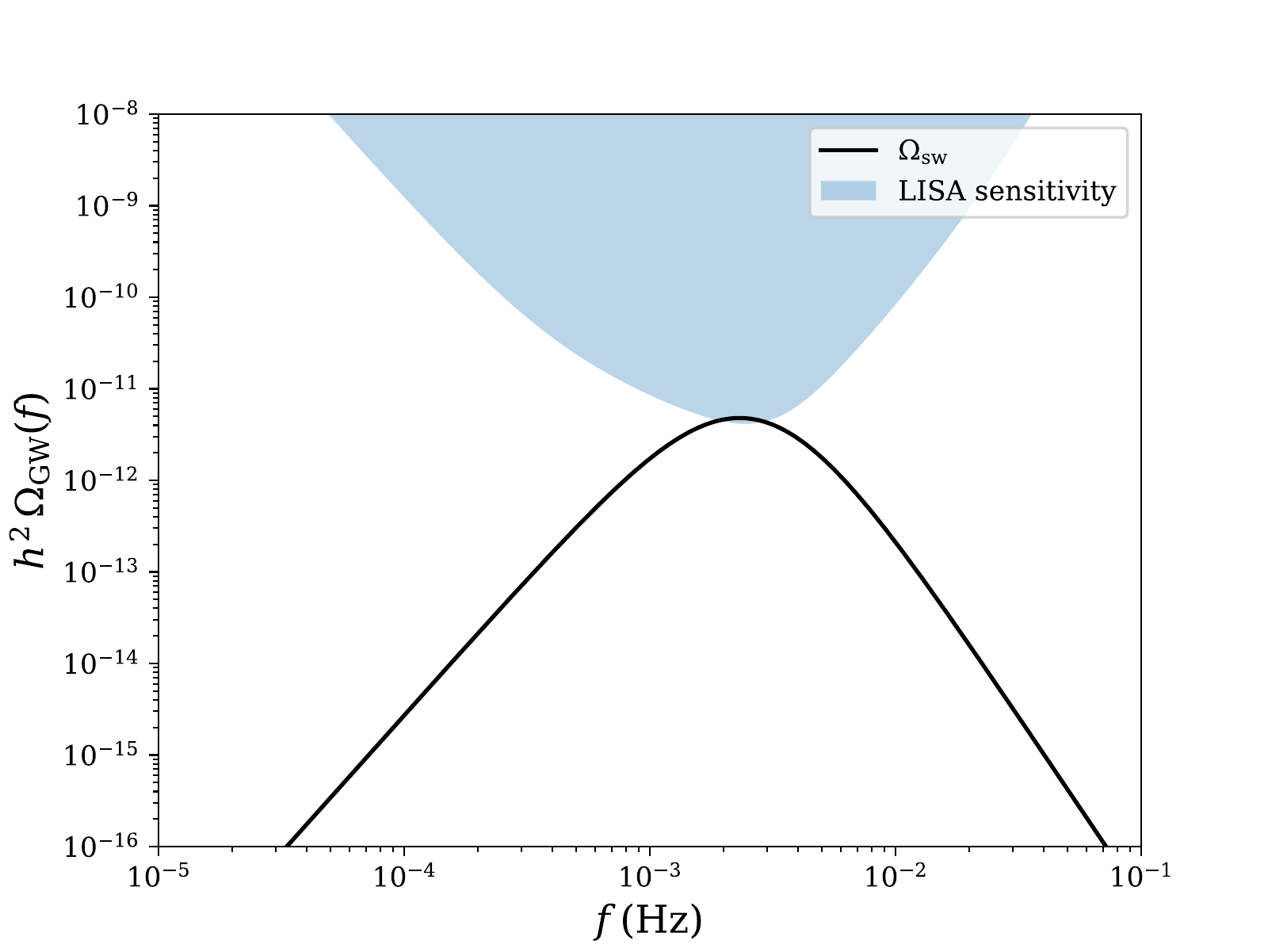}}
    \caption{Typical GW emission in extensions of the Standard Model $(g_*=130)$ in a FOPT with (left) $\alpha=0.6, v_b=0.9,$ $\beta/H_*=100;$  (right) $\alpha=0.2, v_b=0.8.$  We have set $T_*=200$ GeV. }
    \label{one}
\end{figure}

\begin{figure}[ht]
    \centering
\raisebox{-0.5\height}{\includegraphics[scale=0.4]{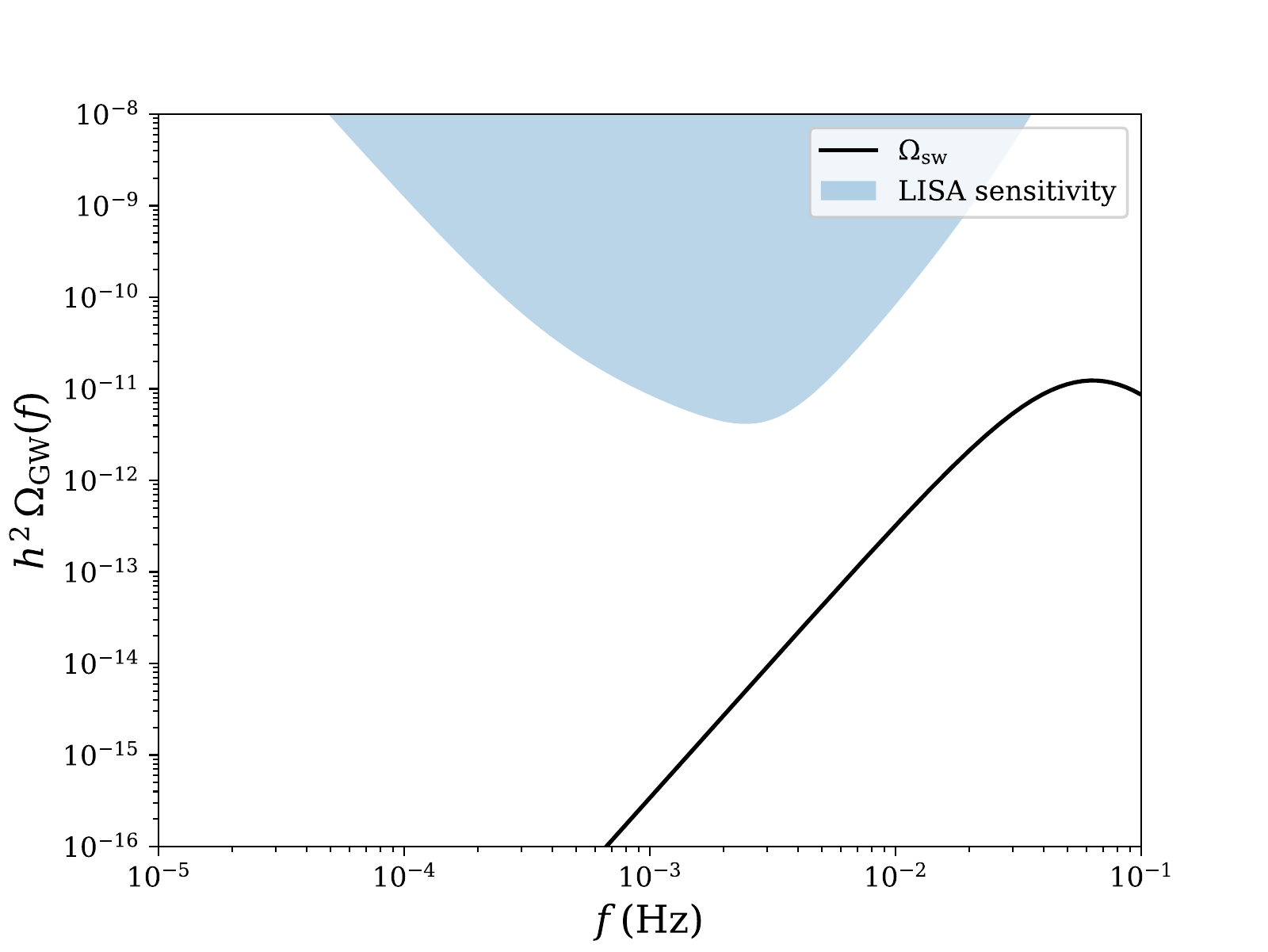}\includegraphics[scale=0.4]{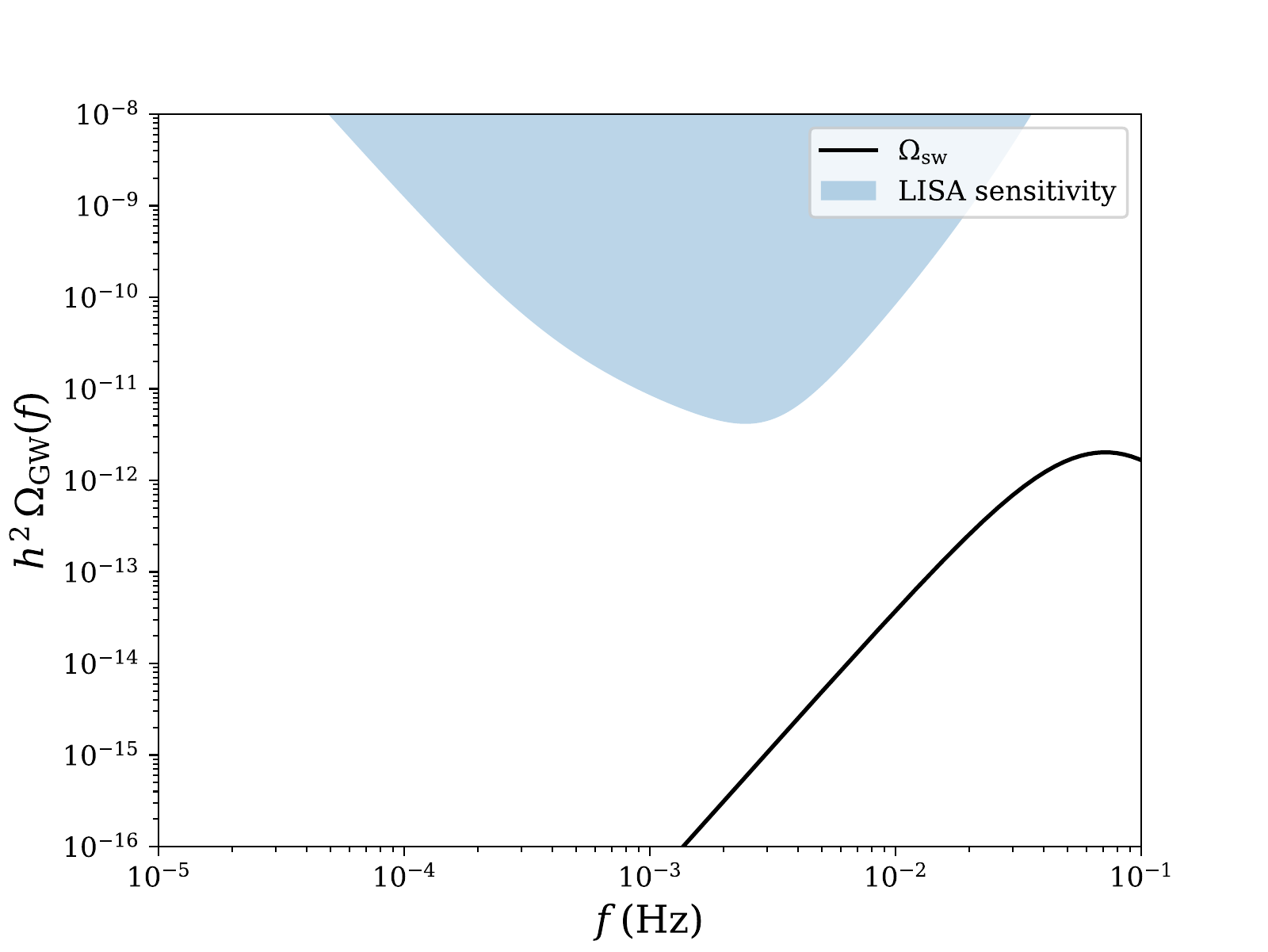}}
    \caption{ GW emission in the quiver model with $g_*=1732$ in a FOPT with (left) $\alpha=0.6, v_b=0.9,$ $\beta/H_*=100;$ (right)  $\alpha=0.2, v_b=0.8$.  We have set $T_*=4000$ GeV. }
    \label{two}
\end{figure}

\begin{figure}[ht]
    \centering
\raisebox{-0.5\height}{\includegraphics[scale=0.4]{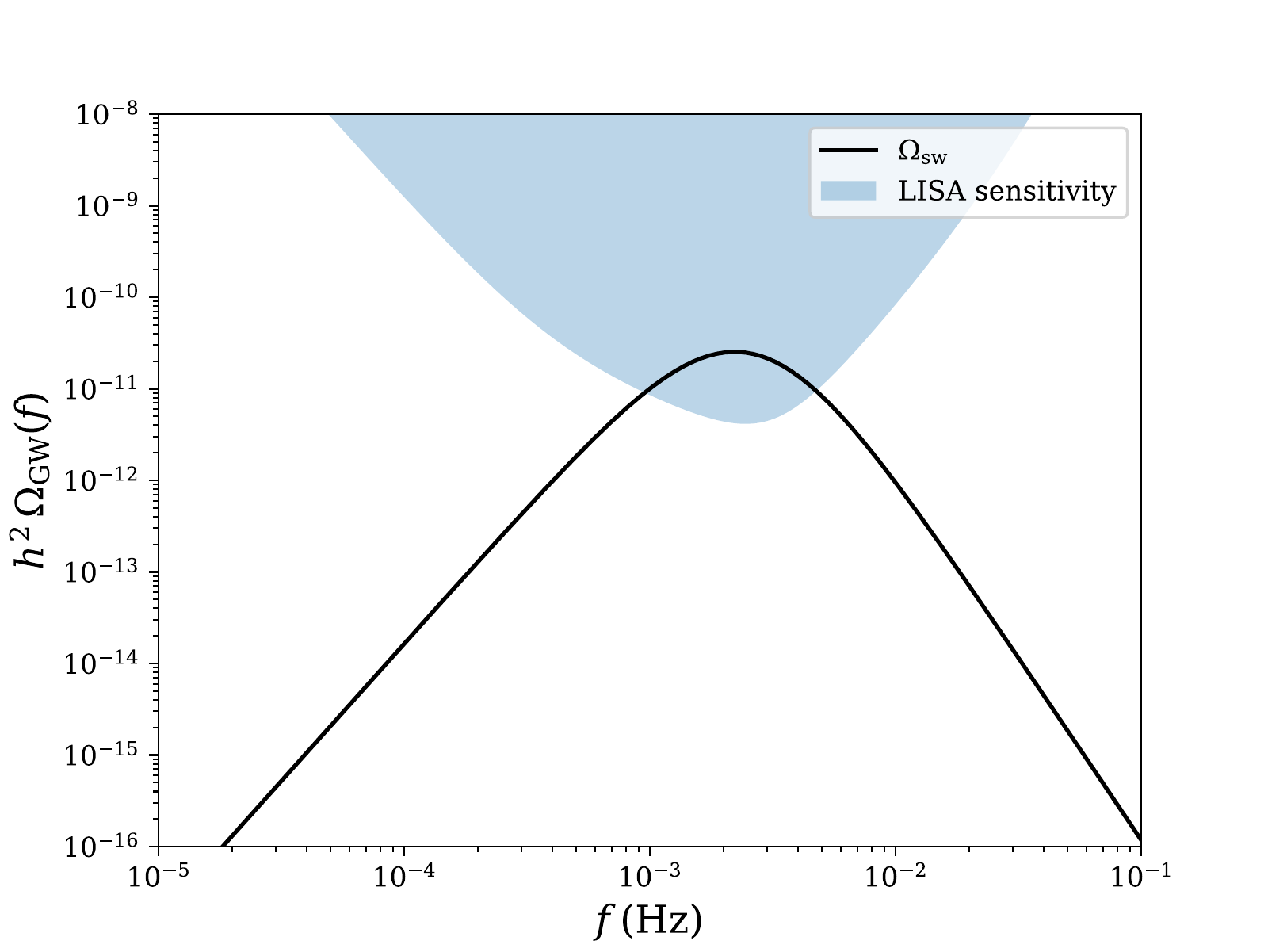}\includegraphics[scale=0.4]{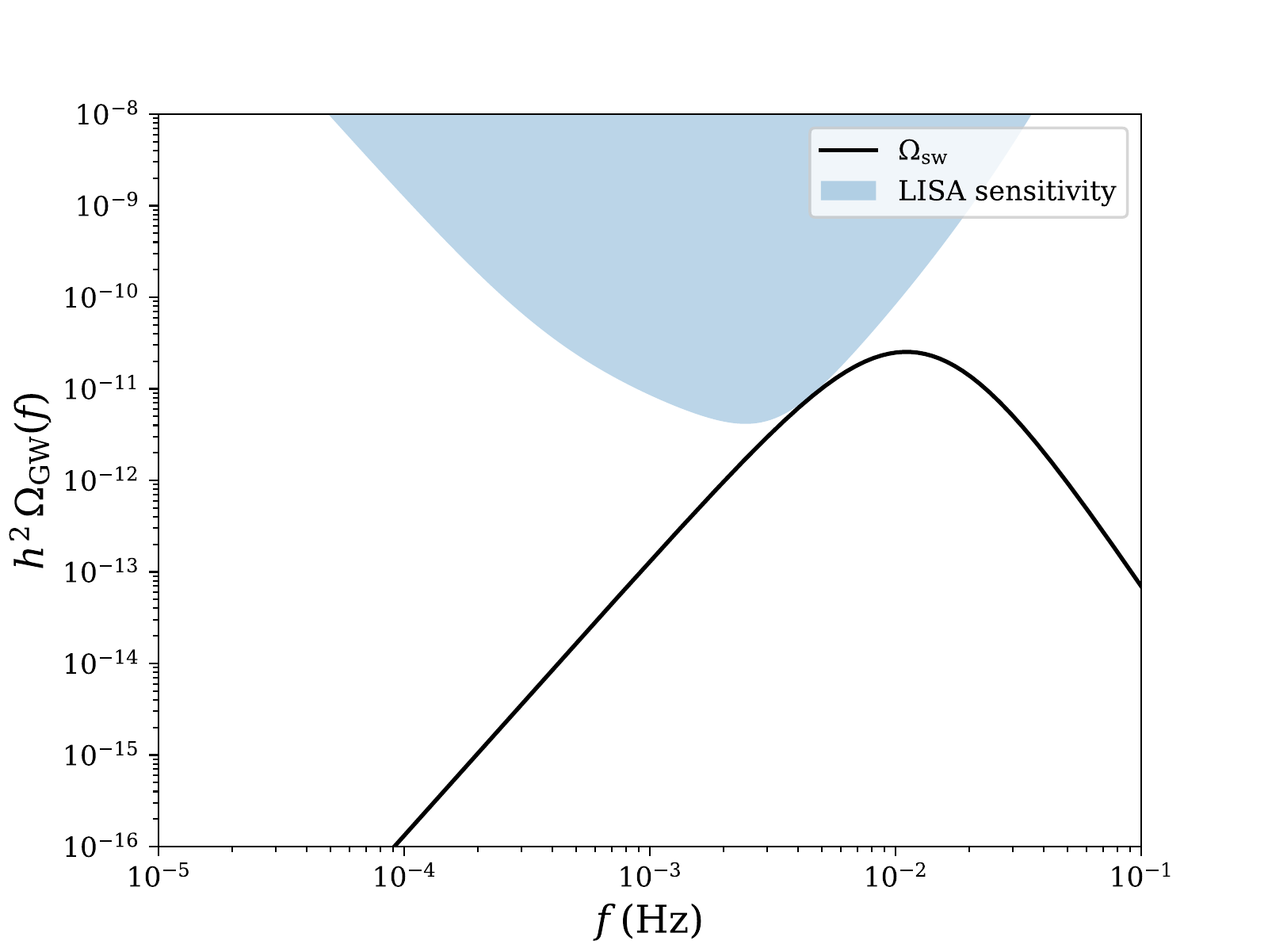}}
    \caption{GW emission in extensions of the Standard Model with (left) $T_*=200$ GeV, $g_*=200$, $\alpha=0.6, v_b=0.9,$ $\beta/H_*=100;$  (right)   $T_*= 1000$ GeV, $g_*=200$ $     \alpha=0.6, v_b=0.9.$ }
    \label{three}
\end{figure}

\begin{figure}[ht]
    \centering
\raisebox{-0.5\height}{\includegraphics[scale=0.4]{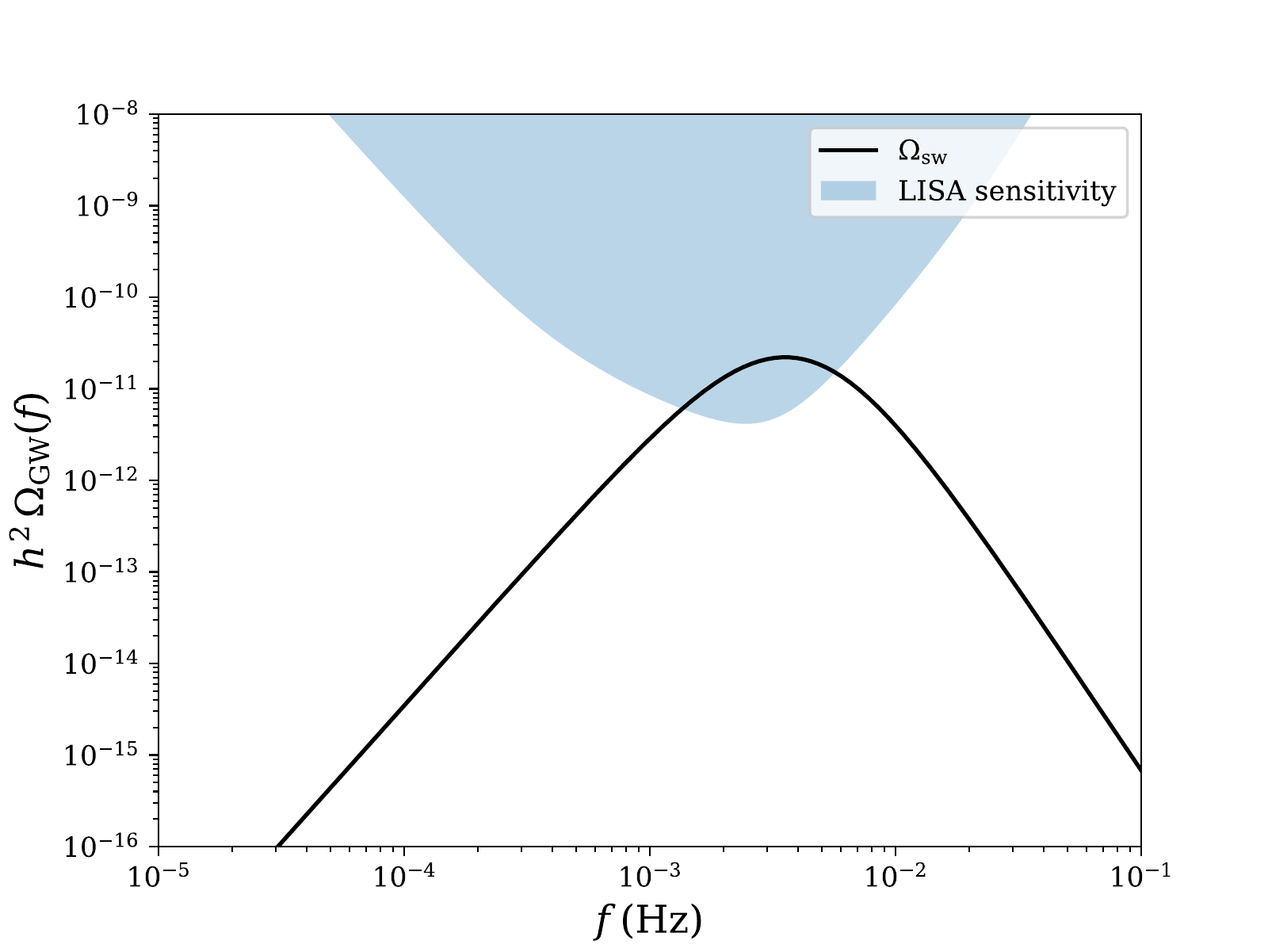}\includegraphics[scale=0.4]{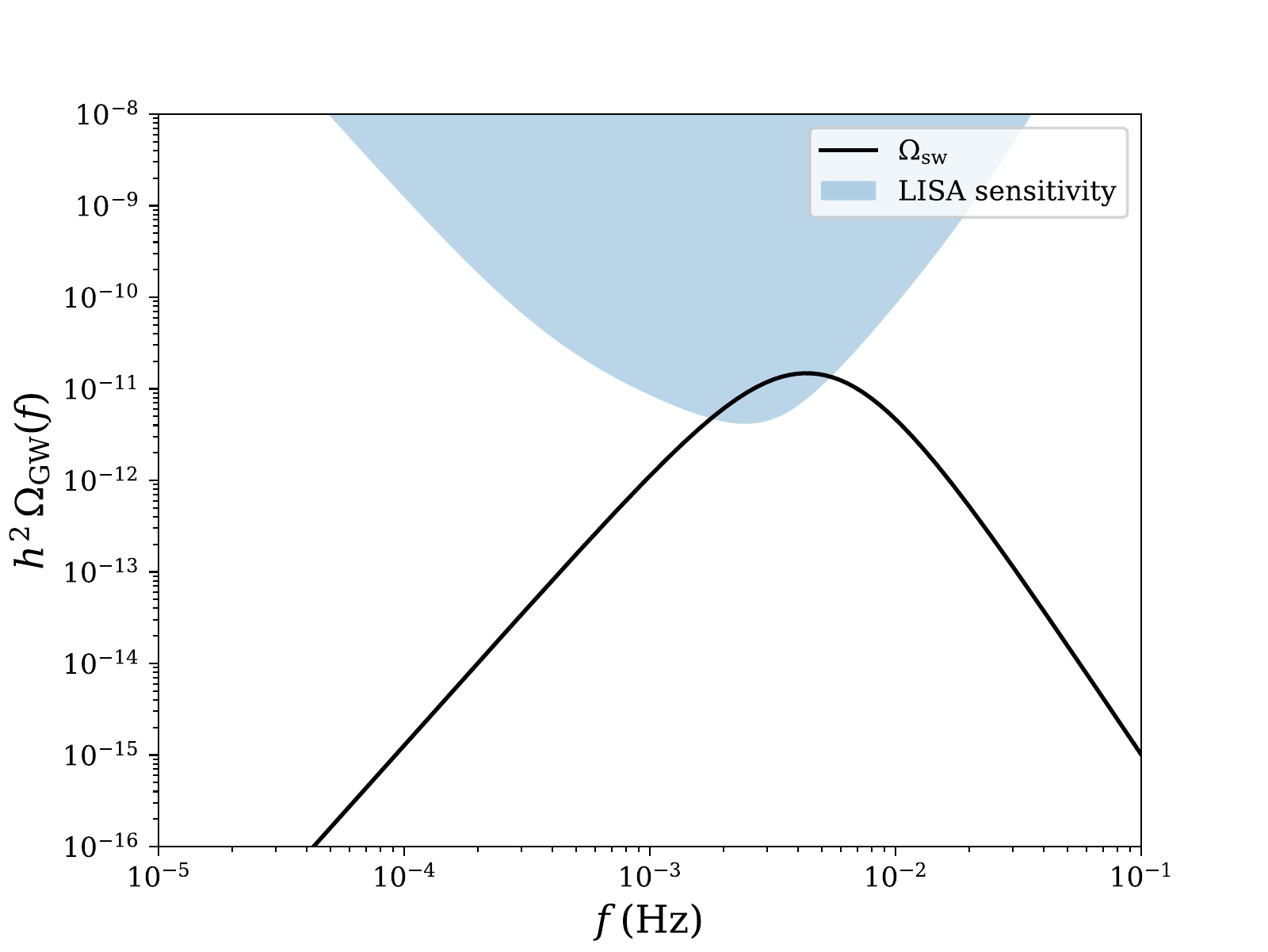}}
    \caption{GW emission in extensions of the Standard Model with (left) $T_*=300$ GeV, $g_*=300$, $\alpha=0.6, v_b=0.9,$ $\beta/H_*=100;$  (right)   $T_*= 300$ GeV, $g_*=1000$ $     \alpha=0.2, v_b=0.8.$ }
    \label{four}
\end{figure}

We show in Tables 1, 2 and 3 the values of the relevant parameters the results for the collisional, sound waves and turbolence contributions to $h_0^2 \Omega$, for $\beta/H^*=100,200$ and 300 and parametric values of $\alpha$, the strength of the transition, varying from 0.6 to 0.8. 
%\begin{itemize}
%\item{\bf Collisional}
%\end{itemize}
In the collisional sector, shown in Table 1, peak frequency emissions are in the range of $10^{-2}$ Hz, with contributions which, for a fixed $\beta/H^*$, are essentially stable, as we vary $\alpha$. The percentile variation of $h_0^2 \Omega_{coll}$ is around $20\%$, for a fixed $\beta/H^*$, as $\alpha$ increases by about $10 \%$ stepwise $(\Delta \alpha=0.5)$. For the same, fixed value of $\alpha$, as we increase $\beta/H^*$ from 100 to 300, the reduction of the gravitational wave emission is about 90 \%.
%\begin{itemize}
%\item{ \bf Sound waves} 
%\end{itemize}

Table 2 summarizes the results for the GW emission due to sound waves in the plasma. In this case, the peaks of the emissions are centered at larger frequencies $(\sim 10^{-1} Hz)$ compared to the collisional contributions, and show, similarly to Table 1, very small variations ($<  1\%$)  as we vary $\alpha$, for a given value of $\beta/H^*$.\\
At a fixed value of the ratio $\beta/H^*$, the GW emission increases in a slightly milder way (by $\sim 10-15 \%$) for $\beta/H^*=100$ as we raise $\alpha$, while it is about $20\%$, as in the previous Table, for $\beta/H^*=200,300$.  As in Table 1, the emission into sound waves, for a given $\alpha$, gets suppressed by $90\%$ in its size as we vary $\beta/H^*$  from 100 to 300.

We show in Table 3 results for the GW emissions due to turbulence. The pattern, also in this case, is similar to those of the previous two cases. The peak frequencies are larger $(\sim 10^{-1})$, by factors of 10 and 100 respect to the sound waves and to the collisional contributions, respectively.  The increase in the GW emission, as we vary $\alpha$, is about 10\%, for a given $\beta/H^*$, while the reduction in the energy of the GW gets reduced about  $60-70 \%$  as we increase $\beta/H^*$ from 100 to 300.  
In all cases, the turbolence contributions are larger than those coming from the collisional and the sound waves at their respective peak frequences, with a factor approximately to 20 for the ratio between $\Omega{sw}\sim 20\times  \Omega_{coll}$ and $\Omega_{turb}\sim 1000-2000\times \Omega_{coll} $.

We can compare our results against the discovery potential of the space detector LISA in few plots using PTPlot \cite{Caprini:2019egz}, assuming in all cases a value of $\beta/H^*=100.$\\
 It is clear that the maximum sensitivity for this proposed experiment is for GW amplitudes with a peak around a few  mHz and 
an energy density of the GW  $h^2 \Omega\sim 10^{-11}$. \\ 
We show 4 plots in Figs. \ref{one} and \ref{two}
which illustrate the difference between the quiver model and typical models characterised by a lower number of degrees of freedom $(\sim 150)$, and a transition temperature comparable with that of the electroweak scale ($\sim 200$ GeV). In Fig. \ref{one} we show results for $h_0^2 \Omega$ for a typical choice of parameters $\alpha=0.6$ and 0.2. In the first case the GW energy density 
follows into the sensitivity region of LISA, while in the second case the curve lapses the region of sensitivity, being tangent to it. 
A similar study can be performed in the quiver model, as shown in Fig. \ref{two}, where the plots show that the increase in temperature by few TeV's increases the frequency of such stochastic background. While the overall energy released as GWs is comparable with the one generated by a transition temperature typical of transitions around electroweak scale, the peak in frequency is shifted upward, and located around $\sim 5\times 10^{-2}$ Hz, beyond the sensitivity of LISA. \\
Obviously, one can investigate the parametric dependence of $h_0^2\Omega$ in a general way, by simply varying the parameters which affect the emission of GWs.
For instance, we can vary $T_*$ from  $200$ GeV  to $1000$ GeV, as well as the number of massless degrees of freedom in $\rho_{rad}$, assuming that in both cases a FOPT is ensured by a sufficiently large value of $\alpha$.\\
The first variation of parameters is shown in Fig. \ref{three}, where on the left we plot the GW emission for the lower case $T_*=200$ GeV and on the right for the 
higher temperature case $T_*=1000$, keeping a value of $g_*=200$, which departs rather modestly from the simplest extensions of the Standard Model compared to the quiver case.
The dependence of such models on the number of degrees of freedom $g_*$ is rather mild, as one can realize from Fig. \ref{four}, where we plot $h_0^2 \Omega$ in models with $T_*=300$ GeV and $g_*= 300$ (left) and 1000 (right). In general an increase in $g_*$ moves the value of $f_{peak}$ slightly towards higher frequency, although it is clear that the dominant effect is related to the drastic change of temperature in the transition, which plays a decisive role in the study of such models.  We note that the final factor in Eq.\eqref{amplitudeSW}  is not included in our plots which were produced using PTPlot software.

\section{Discussion}

\noindent
About forty years ago, around 1980, it appeared likely that minimal $SU(5)$ grand unification theory
\cite{Georgi:1974sy} \cite{Frampton:1979cw}
would agree with experiment and proton decay would soon be
observed, with a lifetime $\sim 10^{30}$ years and with the decay
modes and branching ratios in agreement with the predictions of
minimal $SU(5)$ GUT theory. If so, it would
have been a huge leap forward by factor of at least a trillion ($10^{12}$)
in energy scale above the electroweak scale $\sim 100$ GeV. \\
Unfortunately for this simplest GUT, the proton lifetime for the predicted dominant decay mode $p\rightarrow e^+ \pi^0$
was found experimentally
to be 100 times too long, now known to be 10,000 times too long, to agree
with minimal SU(5).
The reason that minimal $SU(5)$ theory failed was surely because of
the desert hypothesis that there exists no new physics in the huge hierarchy
between the weak scale and the putative GUT scale. \\
\noindent
In the present paper, therefore, we have avoided
this desert hypothesis by employing a quiver GUT which makes no assumption 
about new physics at scales above $4$ TeV, except that the theory
is expected to become conformally invariant up to much higher scales. 
Proton decay is absent at tree level because of the quiver inspired
assignments of the quarks and leptons. If we assume that the breaking of conformal symmetry is characterised by a FOPT at a relatively small scale, in this scenario one should consider the production of gravitational waves in a frequency interval ($10^{-3}-10^{-1}$ Hz) which is in the range of proposed recent experiments. We should also mention that direct simulations \cite{Pol:2019yex} may give the opportunity to improve systematically on previous approximations, especially for what concerns the contribution of $f_{turb}$ to the GEW emissions. \\
  We have suggested that a phase transition in the early universe,
expected by the $SU(3)^{12}$ quiver GUT theory described
in this article, could source GWs in the mHz region, but slightly too large in frequency to be detectable by the forthcoming LISA gravitational wave detector, both for a three and a seven year run of this experiment. However, the wide array of experiments proposed in the future may be able to detect or exclude models with larger transition temperatures respect to those taken into account in the past. \\
 Such models are characterised by a rather large number of massless degrees of freedom compared to the Standard Model or other simpler models, such as the 2-Higgs doublet model, which modify minimally the Standard Model and allow a FOPT to take place rather close to the electroweak scale. In the quiver model that we have discussed, the larger transition temperature $T_*$ and the larger number of degrees of freedom, present a new challenge for their detection both at theoretical and at experimental level.\\

 \vspace{1cm}

\centerline{\bf Acknowledgements}
This work is partly supported by INFN, Iniziativa Specifica QFT-HEP. C.C. thanks M.M. Maglio, D. Theofilopoulos and L. Delle Rose for discussions.

\providecommand{\href}[2]{#2}\begingroup\raggedright\endgroup

\end{document}